\documentclass[10pt]{iopart}

\usepackage{graphicx}
\usepackage{color}
\usepackage{bm}
\usepackage{cite}
\usepackage{multirow}
\usepackage{hyperref}

\def\C60{${\rm C}_{60}$}
\def\VC60{$V_{{\rm C}_{60}^{3-}}$}
\def\A3C60{$A_3{\rm C}_{60}$}
\def\K3C60{${\rm K}_3{\rm C}_{60}$}
\def\Rb3C60{${\rm Rb}_3{\rm C}_{60}$}
\def\Cs3C60{${\rm Cs}_3{\rm C}_{60}$}
\def\Tc{T_{\rm c}}

\begin{document}

%\title[]{Exotic $s$-wave superconductivity in alkali-doped fullerides}
\topical{Exotic $s$-wave superconductivity in alkali-doped fullerides}

\author{Yusuke Nomura$^1$\footnote{Present adress: Centre de Physique Th\'eorique, \'Ecole Polytechnique, CNRS, Universit\'e Paris-Saclay, F-91128 Palaiseau, France.}, Shiro Sakai$^2$,
Massimo Capone$^3$, and Ryotaro Arita$^{2,4}$}

\address{$^1$Department of Applied Physics, University of Tokyo, Hongo, Bunkyo-ku, Tokyo 113-8656,
Japan}
\address{$^2$Center for Emergent Matter Science (CEMS), RIKEN, 2-1, Hirosawa, Wako, Saitama
351-0198, Japan}
\address{$^3$International School for Advanced Studies (SISSA) and Consiglio
Nazionale delle Ricerche-Istituto Officina dei Materiali (CNR-IOM) Democritos National
Simulation Center, Via Bonomea 265, I-34136 Trieste, Italy}
\address{$^4$Japan Science and Technology
Agency (JST) ERATO Isobe Degenerate $\pi$-Integration Project, Advanced Institute for
Materials Research (AIMR), Tohoku University, 2-1-1 Katahira, Aoba-ku, Sendai 980-8577,
Japan.}
\ead{yusuke.nomura@riken.jp}

\vspace{10pt}
\begin{indented}
\item[]\today
\end{indented}

\begin{abstract}
Alkali-doped fullerides (\A3C60 with $A=$ K, Rb, Cs) show a surprising phase diagram, in which 
a high transition-temperature ($\Tc$) $s$-wave superconducting state emerges 
next to a Mott insulating phase as a function of the lattice spacing.
This is in contrast with the common belief 
that Mott physics and phonon-driven $s$-wave superconductivity 
are incompatible, raising
a fundamental question on the mechanism of the high-$\Tc$ superconductivity.  
This article reviews recent {\it ab initio} 
calculations, which have succeeded in reproducing comprehensively the experimental phase diagram with high accuracy and
elucidated an unusual cooperation between the electron-phonon coupling and the electron-electron interactions leading to 
Mott localization to realize an unconventional $s$-wave superconductivity in the alkali-doped fullerides. 
A driving force behind the exotic physics is 
unusual intramolecular interactions, characterized by the coexistence of
 a strongly repulsive 
Coulomb interaction and a small effectively negative exchange interaction.
 This is realized by a subtle energy balance between the coupling with the Jahn-Teller phonons and Hund's coupling within the \C60 molecule.
The unusual form of the interaction leads to a formation of pairs of up- and down-spin electrons on the molecules, 
which enables the $s$-wave pairing.   
The emergent superconductivity crucially relies on the presence of the Jahn-Teller phonons, but surprisingly benefits from the strong correlations because
 the correlations suppress the kinetic energy of the electrons and help the formation of the electron pairs, in agreement with previous model calculations.
This confirms that the alkali-doped fullerides are a new type of unconventional superconductors, where the unusual synergy between the phonons and Coulomb interactions drives the 
high-$\Tc$ superconductivity. 
\end{abstract}

\vspace{1pc}
\noindent{\it Keywords}: Unconventional superconductivity,  alkali-doped fullerides, electron correlations, electron-phonon interactions
\maketitle
\contentsline {section}{\numberline {1}Introduction}{3}{section.1}
\contentsline {subsection}{\numberline {1.1}Crystal structure}{3}{subsection.1.1}
\contentsline {subsection}{\numberline {1.2}Phase diagram}{3}{subsection.1.2}
\contentsline {subsection}{\numberline {1.3}Evidences of phonons}{3}{subsection.1.3}
\contentsline {subsection}{\numberline {1.4}Importance of electron correlations}{4}{subsection.1.4}
\contentsline {subsection}{\numberline {1.5}Aim and outline of this article}{5}{subsection.1.5}
\contentsline {section}{\numberline {2}Methods: DFT+DMFT with including phonon degrees of freedom}{5}{section.2}
\contentsline {section}{\numberline {3}Electronic structure of the fullerides}{6}{section.3}
\contentsline {section}{\numberline {4}Unusual intramolecular interactions}{7}{section.4}
\contentsline {subsection}{\numberline {4.1}Coulomb interactions}{8}{subsection.4.1}
\contentsline {subsubsection}{\numberline {4.1.1}Formulation}{8}{subsubsection.4.1.1}
\contentsline {subsubsection}{\numberline {4.1.2}Results}{8}{subsubsection.4.1.2}
\contentsline {subsection}{\numberline {4.2}Phonon-mediated interactions}{9}{subsection.4.2}
\contentsline {subsubsection}{\numberline {4.2.1}Formulation}{9}{subsubsection.4.2.1}
\contentsline {subsubsection}{\numberline {4.2.2}Electron-phonon coupling}{9}{subsubsection.4.2.2}
\contentsline {subsubsection}{\numberline {4.2.3}Phonon frequencies}{9}{subsubsection.4.2.3}
\contentsline {subsubsection}{\numberline {4.2.4}Results}{10}{subsubsection.4.2.4}
\contentsline {subsection}{\numberline {4.3}Effective interaction between electrons: repulsive Hubbard and negative exchange interactions}{11}{subsection.4.3}
\contentsline {section}{\numberline {5}Theoretical calculation of phase diagram and $T_{\rm c}$ from first principles}{11}{section.5}
\contentsline {subsection}{\numberline {5.1}Theoretical phase diagram}{11}{subsection.5.1}
\contentsline {subsection}{\numberline {5.2}Comparison between theory and experiment}{12}{subsection.5.2}
\contentsline {section}{\numberline {6}Property of metal-insulator transition}{12}{section.6}
\contentsline {subsection}{\numberline {6.1}Single-particle spectral function}{12}{subsection.6.1}
\contentsline {subsection}{\numberline {6.2}Unusual behaviors across the Mott transition}{13}{subsection.6.2}
\contentsline {section}{\numberline {7}Superconducting mechanism}{14}{section.7}
\contentsline {section}{\numberline {8}Conclusion and future perspective}{15}{section.8}
\contentsline {subsection}{\numberline {8.1}Summary of the review}{15}{subsection.8.1}
\contentsline {subsection}{\numberline {8.2}Future perspective}{16}{subsection.8.2}
\contentsline {subsubsection}{\numberline {8.2.1}Study on A15 ${\rm Cs}_3{\rm C}_{60}$}{16}{subsubsection.8.2.1}
\contentsline {subsubsection}{\numberline {8.2.2}Light-induced superconducting-like state in ${\rm K}_3{\rm C}_{60}$}{16}{subsubsection.8.2.2}
\contentsline {section}{\numberline {A}Effect of intermolecular Coulomb interactions}{16}{appendix.A}
\ioptwocol

\section{Introduction}

The alkali-doped fullerides with the composition of \A3C60 ($A=$ K, Rb, Cs) show the highest transition temperature ($\sim$ 40 K) among the molecular superconductors~\cite{C60_super1,PhysRevLett.66.2830,Holczer24051991,tanigaki_nature,fleming_nature,Palstra1995327,A15_CsC60nmat,Takabayashi20032009,PhysRevLett.104.256402,fcc_CsC60,Zadike1500059}. 
The superconductivity was first discovered in \K3C60 by  Hebard {\it et al.} in 1991~\cite{C60_super1}.
Since then, 
%there has been much effort 
much effort has been exerted to understand the mechanism of the
fascinating superconductivity~\cite{Hebard_phys_today,RevModPhys.68.855,RevModPhys.69.575,0034-4885-64-5-202,0953-8984-15-13-202,Gun_book,RevModPhys.81.943}.
Among various works, in this review, we mainly focus on the recent theoretical understanding on the \C60 superconductors~\cite{nomura_science_advances,nomura_cDFPT,PhysRevB.85.155452}. 

\subsection{Crystal structure}
\begin{figure}[tbp]
\vspace{0cm}
\begin{center}
\includegraphics[width=0.55\textwidth]{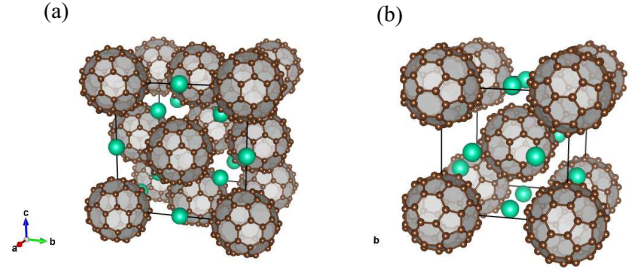}
\caption{Crystal structures of (a) fcc \A3C60 ($A$=K, Rb, Cs) and (b) A15 \Cs3C60, drawn by VESTA~\cite{Momma:db5098}. 
%In the figure for the fcc structure, the experimentally observed disorder in the orientations of the \C60 molecules (merohedral disorder)~\cite{nature_K3C60structure,PhysRevB.45.543,PhysRevB.51.5973} is ignored. 
The orientational disorder of the \C60 molecules (merohedral disorder)~\cite{nature_K3C60structure,PhysRevB.45.543,PhysRevB.51.5973}, which exists in fcc \A3C60, is neglected in (a). 
} 
\label{Fig_structure}
\end{center}
\end{figure}

Figure~\ref{Fig_structure} shows the crystal structure of the \A3C60 systems.   
\K3C60 and \Rb3C60 have the 
face-centered-cubic (fcc) structure~\cite{nature_K3C60structure,PhysRevB.45.543,Zhou19921373}. 
\Cs3C60 %shows, depending on
%the experimental conditions, both fcc and A15 \Cs3C60 
can be synthesized into either an fcc or an A15 structure~\cite{A15_CsC60nmat,PhysRevLett.104.256402,fcc_CsC60}. 
In the fcc structure, the buckyballs are located at the fcc positions and the alkali atoms are intercalated in between the \C60 molecules. 
%The A15 structure is based on the bcc lattice; the positions of the molecules form the bcc lattice, however, the A15 structure contains two molecules in a unit cell since the orientations of the two \C60 molecules are different from each other. 
In the A15 structure, the \C60 molecules are located on the body-centered-cubic (bcc) lattice, while a unit cell contains two \C60 molecules with different orientations.
In both structures,
the intercalated alkali atoms donate electrons to the fullerene bands turning the semiconducting
%into the system, 
%which induces a metallic behavior (the undoped \C60 solids are semiconductors)
%changing the semiconducting property of the 
undoped \C60 solid into a metal~\cite{C60_metal}.   

\subsection{Phase diagram}
\begin{figure}[tbp]
\vspace{0cm}
\begin{center}
\includegraphics[width=0.44\textwidth]{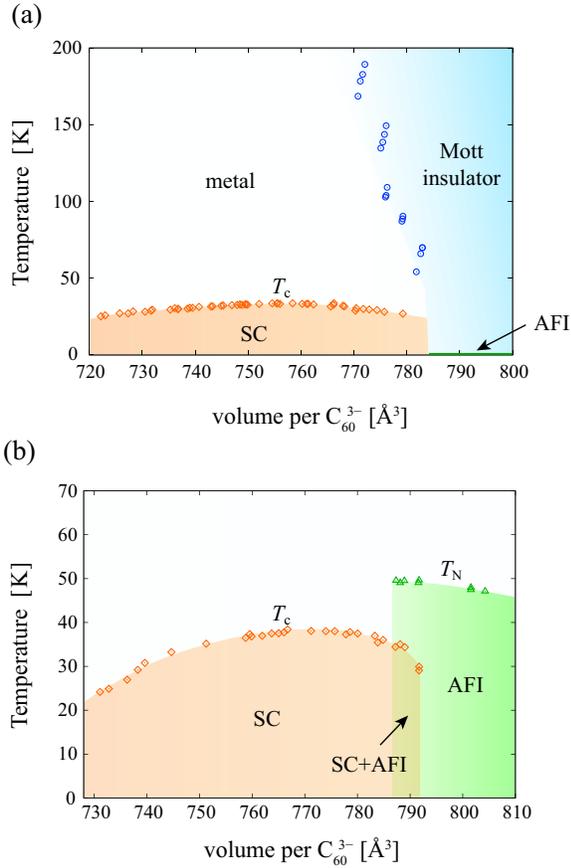}
\caption{
Experimental phase diagram for (a) fcc \A3C60 ($A$=K, Rb, Cs) and (b) A15 \Cs3C60.
The data points are taken from Refs.~\cite{Zadike1500059} and \cite{Takabayashi20032009} for (a) and (b), respectively. 
SC and AFI denote the superconducting phase and the anti-ferromagnetic insulator, respectively. 
$\Tc$ (open diamonds) and $T_{\rm N}$ (open triangles) are the superconducting transition temperature and the N\'eel temperature, respectively. 
The open circles indicate the crossover between the metal and the insulator. 
In the panel (a), the data points for \K3C60, \Rb3C60, \Cs3C60, and the fullerides with mixed alkali composition such as Rb$_2$CsC$_{60}$ are included. 
In the phase diagram of fcc \A3C60 [panel(a)], while not shown explicitly, there is a region close to the Mott insulating phase, which is dubbed ``Jahn-Teller metal" in Ref.~\cite{Zadike1500059} (see the main text for detail). 
The volume per C$_{60}^{3-}$ anion is given by $V_{{\rm C}_{60}^{3-}} = a^3/4$ for the fcc systems and $V_{{\rm C}_{60}^{3-}} = a^3/2$ for the A15 systems, where $a$ is the lattice constant.
} 
\label{Fig_phase}
\end{center}
\end{figure}

Figure~\ref{Fig_phase} shows the %latest 
most refined experimental phase diagrams for the (a) fcc and (b) A15 structures~\cite{A15_CsC60nmat,Takabayashi20032009,PhysRevLett.104.256402,fcc_CsC60,0295-5075-94-3-37007}.  
In both cases, the horizontal axis is labelled by the  volume \VC60 occupied per ${\rm C}_{60}^{3-}$ anion in the actual solid structure. 
In the fcc case, \VC60 is controlled by physical and/or chemical pressure, where the latter is induced by changing the alkali species, with larger ions leading to larger lattice spacing between molecules. 
The phase diagram for the A15 structure was
obtained by applying the physical pressure to \Cs3C60 solid~\cite{Takabayashi20032009}. 
%In the phase diagrams, there exists the superconducting phase. 

Both phase diagrams show a superconducting phase with a high critical temperature, whose maximum reaches $\sim 35$ K and $\sim 38$ K for the fcc and A15 systems, respectively.
% with a dome-shaped behavior as a function of  \VC60 which reminds the 
%typical behavior of cuprate superconductors and other superconducting families. 
Interestingly, 
%the shapes of the $\Tc$ dome are quite similar in fcc and A15 phase diagram.
the fcc and A15 phase diagrams share a similar shape of the $\Tc$ dome 
 as a function of  \VC60 despite the lack of magnetic ordering in the Mott state for the frustrated fcc lattice.
The symmetry of the superconducting order parameter was found to be of $s$-wave by 
various different experimental techniques~\cite{gap_Rb3C60,ZHANG13121991,PhysRevLett.68.1912,AubanSenzier19933027,doi:10.1143/JPSJ.63.1670,PhysRevLett.70.3987,gap_Rb3C60_2,Nature369541,PhysRevLett.77.4082,PhysRevB.50.16566,PhysRevLett.85.1970} as in the whole
family of alkali-doped fullerides.
%e.g. the STM (scanning tunneling microscope)~\cite{gap_Rb3C60,ZHANG13121991}, the NMR (nuclear magnetic resonance)~\cite{PhysRevLett.68.1912,AubanSenzier19933027,doi:10.1143/JPSJ.63.1670}, 
%the $\mu$SR (muon spin rotation/relaxation/resonance)~\cite{PhysRevLett.70.3987},
%the infrared spectroscopy~\cite{gap_Rb3C60_2,Nature369541,PhysRevLett.77.4082}, 
% the tunneling measurement~\cite{PhysRevLett.77.4082}, and the photoemission spectroscopy~\cite{PhysRevB.50.16566,PhysRevLett.85.1970}. 
Also the suppression of the spin fluctuation in the superconducting state suggests a singlet pairing~\cite{PhysRevLett.74.1649}.

What makes the system 
%even 
more remarkable is the existence of the Mott insulating phase next to the $s$-wave superconducting phase~\cite{A15_CsC60nmat,Takabayashi20032009,PhysRevLett.104.256402,fcc_CsC60}. 
This adjacency of the superconductivity and the Mott insulator is reminiscent of the cuprates~\cite{discover_Cu},
where the symmetry of the order parameter is established to be $d$-wave~\cite{RevModPhys.72.969,RevModPhys.75.473}. 
In fullerides, this proximity is even more surprising 
because the $s$-wave superconductivity is believed to be fragile to the strong correlations, as opposed to the $d$-wave symmetry.

In the Mott phase, at low temperature,  
the antiferromagnetism has been observed in both fcc and A15 systems~\cite{Takabayashi20032009,fcc_CsC60,0295-5075-94-3-37007}. 
However, the N\'eel 
temperatures $T_{\rm N}$ are very different between the two systems.
In the A15 systems, which are less frustrated than the fcc systems, the transition into the antiferromagnetic long range order with the wave vector 
${\bf q} = (\frac{1}{2}, \frac{1}{2}, \frac{1}{2})$ occurs at around 46 K~\cite{PhysRevB.80.195424}. 
In the fcc case, the magnetic instability is drastically suppressed 
%($T_{\rm N} \sim$ 2 K). 
to $T_{\rm N} \sim$ 2 K.
Even below $T_{\rm N}$, the magnetism is not purely long-range ordered: 
A specific heat measurement suggests the coexistence of the glass-like disordered magnetism and the antiferromagnetic order below $T_{\rm N}$~\cite{PhysRevB.90.014413}. 
%The specific heat measurement suggests the coexistence of 
%the glass-like disordered magnetism and the antiferromagnetic order below $T_{\rm N}$~\cite{PhysRevB.90.014413}.
%This suppression is ascribed to the geometrical frustration 
%and of the disorder in the superexchange type spin-spin interactions~\cite{PhysRevB.90.014413}. 
The suppression of magnetism and the complicated magnetic structure in the fcc systems 
are ascribed to the geometrical frustration and to disorder in the superexchange interactions driving the coupling between localized spins~\cite{PhysRevB.90.014413}.

\subsection{Evidences of phonons}
In 1990's (well before the discovery of the Mott phase and magnetism in \Cs3C60 in 2008), the pairing mechanism had been often discussed based on the conventional phonon mechanism~\cite{RevModPhys.69.575,VARMA15111991,PhysRevLett.68.526,PhysRevB.45.2597,PhysRevB.45.5114}. 
Indeed, within the Migdal-Eliashberg theory~\cite{Schrieffer_book}, the theoretically and experimentally estimated electron-phonon coupling constant $\lambda  \sim 0.5$-1 
~\cite{VARMA15111991,PhysRevLett.68.526,PhysRevB.45.2597,PhysRevB.45.5114,PhysRevB.46.1265,PhysRevB.48.661,PhysRevB.48.7651,doi:10.1080/13642810110062663,Breda1998350,PhysRevB.65.220508,PhysRevB.81.073106,PhysRevB.84.155104,PhysRevB.88.054510,PhysRevLett.81.697,PhysRevLett.74.1875,PhysRevB.53.655,PhysRevB.57.608,PhysRevLett.72.4121,Marsiglio1998172,PhysRevB.77.115445,PhysRevB.82.245409}
 and the high phonon frequency $\sim 0.1$ eV~\cite{VARMA15111991,PhysRevLett.68.526,PhysRevB.45.2597,PhysRevB.45.5114,PhysRevB.46.1265,PhysRevB.48.661,PhysRevB.48.7651,doi:10.1080/13642810110062663,Breda1998350,PhysRevB.65.220508,PhysRevB.81.073106,PhysRevB.84.155104,PhysRevB.88.054510,PhysRevLett.74.1875,PhysRevB.53.655,PhysRevB.57.608,PhysRevB.77.115445,PhysRevB.82.245409,Bethune1991181,PhysRevB.45.10838,PhysRevB.51.5805,Quong1993535,:/content/aip/journal/jcp/100/11/10.1063/1.466753}
 led to the prediction of a transition temperature comparable to the experimental value 
 under the assumption that the Coulomb repulsion is reasonably weak. 
The phonon mechanism seemed
 consistent with the experimental observation of the Hebel-Slichter peak~\cite{doi:10.1143/JPSJ.63.1670,PhysRevLett.70.3987} and the isotope effect with the exponent $\sim 0.2$-0.3~\cite{doi:10.1021/ja00034a072,PhysRevLett.83.404}, 
too.\footnote{We quote the exponent measured for 99 \% $^{13}$C-rich samples. The exponents obtained for the samples with incomplete substitution range from $\sim$ 0.3 to $\sim 2.1$~\cite{AubanSenzier19933027,Zakhidov1992355,isotope_Rb3C60,PhysRevLett.68.1058}.} 
The positive correlation between $\Tc$ and the lattice constant~\cite{fleming_nature,SPARN28061991,Schirber1991137,ZHOU14021992,PhysRevLett.68.1228} also 
supported this scenario: 
When the lattice is expanded, the bandwidth decreases and hence  
the density of states (DOS) at the Fermi level increases.
According to the Bardeen-Cooper-Schrieffer (BCS) theory, it leads to the increase of $\Tc$~\cite{tanigaki_nature,fleming_nature}. 

\subsection{Importance of electron correlations}
On the other hand, the importance of the electron correlations has been argued since the early stage of the study. 
Auger spectroscopy measurements for the undoped \C60 solid  
lead to an estimate of $\sim 1.6$ eV for the effective intramolecular Coulomb interaction~\cite{PhysRevLett.68.3924}, which is bigger than the typical bandwidth of the low-energy bands $\sim 0.5$ eV.
Furthermore, it has been argued that the retardation effect might be inefficient~\cite{PhysRevLett.69.957,Gun_pseudo,PhysRevLett.69.212,Gun_book}; because the electronic bands of the fullerides are distributed sparsely in energy, the density of states does not spread continuously any more~\cite{ERWIN08111991,PhysRevB.44.11536,PhysRevLett.101.136404} (Sec.~\ref{sec_ele}). Based on this fact, the typical electronic energy scale has been argued to be given by the bandwidth of the low-energy $t_{1u}$ bands ($\sim 0.5$ eV), which are energetically isolated from the other bands.  
This leads to a large Coulomb pseudopotential, which 
%strongly 
challenges the conventional pairing mechanism~\cite{PhysRevLett.69.212}. 
There are also suggestions of purely electronic mechanism based on the resonating valence bond scenario~\cite{Baskaran,0295-5075-16-8-008,CHAKRAVARTY15111991}. 
Several works have taken into account both the electron correlations and the electron-phonon interactions to explain the superconductivity~\cite{RevModPhys.81.943,Takada19931779,PhysRevLett.86.5361,Capone28062002,PhysRevLett.93.047001,PhysRevLett.90.167006}.

The discovery of the Mott insulating phase in \Cs3C60 strikingly confirmed the strength and the relevance of the electron-electron correlations~\cite{A15_CsC60nmat,Takabayashi20032009,PhysRevLett.104.256402,fcc_CsC60,0295-5075-94-3-37007}. 
Through the metal-insulator transition, there is no structural transition~\cite{A15_CsC60nmat,Takabayashi20032009,fcc_CsC60}.
Therefore, the alkali-doped fullerides provide a unique playground to study the $s$-wave superconductivity under the strong correlation. 

This discovery has triggered both the experimental and theoretical studies~\cite{Zadike1500059,nomura_science_advances,nomura_cDFPT,PhysRevB.85.155452,0295-5075-94-3-37007,IR_ncom,1742-6596-428-1-012002,C4SC00670D,doi:10.7566/JPSJ.82.014709,PhysRevLett.112.066401,PhysRevB.86.085109,doi:10.7566/JPSJ.82.054713,PhysRevLett.109.166404,PhysRevLett.101.136404,Yamazaki_2,PhysRevLett.111.056401,PhysRevB.91.035109,PhysRevB.88.165430,Durajski_1,Durajski_2,kivelson_paper,Baldassarre_srep_2015}. 
As a result, 
%an unusual nature both in 
various unusual properties in both metallic and insulating phases have been revealed. 
In the Mott phase, the size of 
the local spin per \C60 molecule was found to be $S = 1/2$ (low-spin state)~\cite{Takabayashi20032009,fcc_CsC60}, not $S=3/2$ (high-spin state) expected from Hund's rule. 
The analysis of the infrared (IR) spectrum revealed the presence of the dynamical Jahn-Teller distortion of the \C60 molecules~\cite{IR_ncom,1742-6596-428-1-012002,Baldassarre_srep_2015}. 
The nuclear magnetic resonance (NMR), which 
%is a slower probe
probes a slower dynamics than the IR spectroscopy, observed a gradual freezing of the Jahn-Teller dynamics as the temperature decreases~\cite{C4SC00670D}. 

The metallic and superconducting states also show interesting behaviors near the metal-insulator transition. 
In the superconducting state,
 NMR measurements observed a deviation of the ratio between the gap($\Delta$) and $\Tc$  ($2\Delta/k_{\rm B}\Tc$) from the BCS value of 3.53 to a 
larger value, while $2\Delta/k_{\rm B}\Tc \sim 3.53$ holds in the region of  small \VC60~\cite{PhysRevLett.112.066401,srep_gap}. 
The spin susceptibility in the normal phase
also shows a larger value than that expected from the smooth extrapolation from 
the values in the small \VC60 region~\cite{0295-5075-94-3-37007}. 
Moreover, an anomalous metallic region has been identified  close to the metal-insulator boundary, which has been dubbed ``Jahn-Teller metal".
In this state,
the IR spectrum is similar to that of the Mott insulating phase~\cite{Zadike1500059}.
This result can be interpreted in terms of a slowing down of the dynamical distortion of the \C60 molecules when  \VC60 increases. When 
the Mott localization is approached, 
the distortion timescale becomes eventually so long that the IR experiment probes the system in a distorted state
on its characteristic timescale. 

\subsection{Aim and outline of this article}
The clear fingerprints of the electron-phonon coupling in the superconducting state and the very existence of the Mott insulating state suggest the importance of considering both electron correlations and phonons for the understanding of the surprising phase diagram. 
In particular, it is a great challenge to understand why the $s$-wave superconductivity is robust against (or even benefits from) the strong correlations. 

Another challenge is an {\it ab initio} calculation of $\Tc$ of strongly-correlated unconventional superconductivity. 
A nonempirical calculation of $\Tc$ is necessary for predicting/designing 
new high-temperature superconductors.  
Even for the conventional superconductors, the $\Tc$ calculation usually relies on 
%the empirical parameter 
empirical parameters
such as the Coulomb pseudopotential~\cite{PhysRev.125.1263}. 
The recent development of the density-functional theory for superconductors (SCDFT) has enabled a $\Tc$ calculation without empirical parameters~\cite{PhysRevLett.60.2430,PhysRevLett.86.2984,PhysRevB.72.024545,PhysRevB.72.024546}. 
By assuming the phonon-mechanism, the SCDFT has succeeded in 
%predicting 
reproducing $\Tc$ of conventional superconductors 
in the accuracy of several tens percent.
However, $\Tc$ of \C60 superconductors is largely underestimated by the phonon-mechanism-based SCDFT~\cite{PhysRevB.88.054510}.  
This is another indirect indication that the alkali-doped fullerides are unconventional superconductors. 
While there have been several attempts to generalize the SCDFT to include e.g. the plasmon~\cite{PhysRevLett.111.057006,doi:10.7566/JPSJ.83.061016,PhysRevB.91.224513} and the spin-fluctuation~\cite{PhysRevB.90.214504,Gross_arxiv} as a pairing glue, 
there exists no established method to calculate $\Tc$ of unconventional superconductors in which strong-correlation effects are important. 
%the current SCDFT needs to assume some mechanism to calculate $\Tc$.  
%In the case of fullerides, the mechanism is not known {\it a priori}. 

%Hence, in
In this review, among the various studies on the \C60 superconductors, 
we mainly focus on the most recent {\it ab initio} studies~\cite{nomura_science_advances,nomura_cDFPT,PhysRevB.85.155452}, 
which aimed at (i) the unified description of the phase diagram including the $s$-wave superconductivity 
and the Mott phase and (ii) the nonempirical calculation of $\Tc$. 
The nonempirical calculations have elucidated that, in the fullerides, the phonon-mediated negative exchange interaction 
surpasses
the positive Hund's coupling and thereby realizes an inverted Hund's rule, as predicted by Capone {\it et al}.~\cite{RevModPhys.81.943,PhysRevLett.86.5361,Capone28062002,PhysRevLett.93.047001}. 
On the other hand, the intramolecular Hubbard-type interaction is strongly repulsive 
because the strong local Coulomb interaction 
%dominates
far exceeds
the attraction mediated by phonons.  
The value of this repulsion is larger than that of the 
%typical
$t_{1u}$ bandwidth, which brings about the Mott physics in the system. 

 By analyzing a realistic low-energy Hamiltonian with the above-mentioned unusual interactions, the theoretical phase diagram was derived without empirical parameters. 
 Remarkably, it shows a good agreement with the experimental phase diagram at a quantitative level. 
 In particular, the calculated $\Tc$'s agree with the experimental data
 %, where the difference is within 10 K.
 within a difference of 10 K.
 It indicates that the scheme employed in Refs.~\cite{nomura_science_advances,nomura_cDFPT,PhysRevB.85.155452}  
 properly captures the essence of the fulleride superconductivity. 
Based on the success of our approach, we argue that the unusual intramolecular interaction is the key to explain the phase diagram in a unified manner:  
It allows electrons to form a pair on the molecules in contrast to the na\"ive expectation that the electron correlations drastically suppress the pair formation~\cite{PhysRevLett.90.167006}.  Surprisingly, the strong correlations are found to even help the formation of the pair.
We show that this leads to a surprising cooperation between the phonons and Coulomb interactions to realize an exotic high-$\Tc$ pairing next to the Mott phase.

The outline of this review is as follows. 
Throughout the review, we mainly focus on the fcc systems. 
In Sec.~\ref{sec_method}, we describe the methods, which were employed in Refs.~\cite{nomura_science_advances,nomura_cDFPT,PhysRevB.85.155452}.
The methods construct, from first principles, a realistic three-band Hamiltonian with including the phonon degrees of freedom from only the information of the crystal structure. 
We solve it accurately by means of 
a many-body method. 
Through the derivation of the realistic Hamiltonian, we discuss the electronic structure of the fullerides (Sec.~\ref{sec_ele}) and the detail of the above-mentioned unusual intramolecular interaction (Sec~\ref{sec_int}).   
The analysis of the derived model follows in Secs.~\ref{sec_phase}, ~\ref{sec_MIT}, and ~\ref{sec_SC}. 
First, we show the theoretical phase diagram in Sec.~\ref{sec_phase}. 
Then, we discuss the properties of the metal-insulator transition in Sec.~\ref{sec_MIT} and finally the superconducting mechanism in Sec.~\ref{sec_SC}. 
We give a summary of the review and future perspectives in Sec.~\ref{sec_summary}.

\section{Methods: DFT+DMFT with including phonon degrees of freedom}
\label{sec_method}
In the case of the alkali-doped fullerides, an accurate description of the intramolecular correlations induced by the Coulomb interactions and the intramolecular vibrations is important for clarifying the underlying physics. 
The main playground of the intramolecular correlations and the intriguing low-energy phenomena is the partially-filled bands near the Fermi level. 
In the fullerides, the LUMO (lowest unoccupied molecular orbital) bands, the $t_{1u}$ bands, are partially filled (see Sec.~\ref{sec_ele}) and they are energetically isolated from the other bands~\cite{ERWIN08111991,PhysRevB.44.11536,PhysRevLett.101.136404}.  
Therefore, the point is how accurately we describe the intramolecular dynamics involving the $t_{1u}$ electrons and the phonons. 

For this purpose, one of the most appropriate schemes would be a combination of the density-functional theory (DFT) and the dynamical mean-field theory (DMFT), so called DFT+DMFT~\cite{RevModPhys.78.865}. 
The DMFT can accurately take into account the local dynamical correlation effect induced by the phonons 
as well as the Coulomb interaction, while it neglects the spatial correlation effect~\cite{RevModPhys.78.865,RevModPhys.68.13,RevModPhys.83.349}. 
It becomes a better approximation as the spatial dimension increases. 
In fcc \A3C60 having a frustrated lattice with the coordination number of 12, the DMFT is expected to give reliable results. 

In order to make a quantitative argument, we need a realistic low-energy Hamiltoinan for the fullerides to be used in the DMFT calculation. 
This can be done by the {\it ab initio} downfolding scheme~\cite{RevModPhys.78.865,doi:10.1143/JPSJ.79.112001,nomura_cDFPT}: It starts from the DFT band structure and constructs an effective Hamiltonian consisting of the low-energy electrons and the phonons, with including the renormalization effect of the high-energy electrons. 
%With this strategy, 
By solving thus-constructed low-energy Hamiltonian with the DMFT,
the strong-correlation effect within the partially-filled low-energy bands such as the Mott physics, 
which cannot be captured by the conventional DFT, is properly taken into account in an {\it ab initio} way~\cite{RevModPhys.78.865,doi:10.1143/JPSJ.79.112001}.  
Here, we use the word ``{\it ab initio}" for calculations without employing empirical parameters.  

In fact, in Refs.~\cite{nomura_science_advances,nomura_cDFPT,PhysRevB.85.155452}, 
we further extended the above DFT+DMFT scheme to perform {\it ab initio} studies on the fullerides.
%to achieve {\it ab initio} studies on the fullerides, 
%a generalized scheme based on the DFT+DMFT was applied. 
%The usual DFT+DMFT often neglects the phonon degrees of freedom and considers the Hubbard-like model. 
%The scheme employed in Refs.~\cite{nomura_science_advances,nomura_cDFPT,PhysRevB.85.155452} explicitly included the phonon degrees of freedom in the Hamiltonian, 
%as well as the low-energy electron degrees of freedom. 
%The resulting low-energy Hamiltonian consists of%The model was 
The outline of this generalized scheme, 
which explicitly considers the phonon degrees of freedom, 
% employed in Refs.~\cite{nomura_science_advances,nomura_cDFPT,PhysRevB.85.155452}
is as follows. 
\begin{enumerate}
\item{{\bf Band structure calculation} (Sec.~\ref{sec_ele}): Perform the DFT calculation for the alkali-doped fullerides and 
obtain the band structure in a global energy scale.}
%Choose the active electron degrees of freedom (partially-filled bands), for which the low-energy Hamiltonian is constructed. In the case of the fullerides, the LUMO (lowest unoccupied molecular orbital) bands become partially filled (see Sec.~\ref{sec_ele_st})}
\item{  $\bm{Ab\  initio}$ {\bf downfolding} (Secs.~\ref{sec_ele} and \ref{sec_int}): Construct a realistic Hamiltonian for the fullerides. The Hamiltonian is defined for the low-energy $t_{1u}$ electrons and the phonons, and is comprised of the electron hopping $\hat{{\mathcal H}}_{\rm el}$, Coulomb interaction $\hat{{\mathcal H}}_{\rm el\mathchar`-el}$, electron-phonon coupling $\hat{{\mathcal H}}_{\rm el\mathchar`-ph}$, and phonon one-body $\hat{{\mathcal H}}_{\rm ph}$ terms: 
\begin{eqnarray}
 \hspace{0.8cm} \hat{{\mathcal H}}   =   \hat{{\mathcal H}}_{\rm el} + 
  \hat{{\mathcal H}}_{\rm el\mathchar`-el} + 
 \hat{{\mathcal H}}_{\rm el\mathchar`-ph} + 
  \hat{{\mathcal H}}_{\rm ph}.  
  \label{H_ele_ph_system}
\end{eqnarray}
All the parameters in the Hamiltonian are derived by the {\it ab initio} downfolding scheme. }
\item{{\bf Analysis of the Hamiltonian} (Secs.~\ref{sec_phase}, \ref{sec_MIT}, and \ref{sec_SC}): Solve the realistic low-energy Hamiltonian in Eq.~(\ref{H_ele_ph_system}) by the extended DMFT (E-DMFT)~\cite{PhysRevB.52.10295,PhysRevLett.77.3391,PhysRevB.66.085120,PhysRevLett.92.196402,PhysRevB.87.125149,PhysRevB.90.195114} to reveal the exotic physics beneath the phase diagram. 
The E-DMFT takes into account the dynamical screening effect of the non-local (intermolecular in the case of the fullerides) Coulomb interactions, on top of the correlations incorporated by the DMFT.\footnote{We note that the self-energy is still momentum independent within the E-DMFT.} } 
\end{enumerate}
With this three-step scheme, the purely theoretical phase diagram for the fcc system was obtained by using only the information of the crystal structure~\cite{nomura_science_advances}. 
We found that the theoretical phase diagram agrees well with the experimental phase diagram in Fig.~\ref{Fig_phase}(a) even quantitatively (Sec.~\ref{sec_phase}). 
The quantitative agreement allows us to 
make a conclusive remark on the mechanism of the high-$\Tc$ $s$-wave superconductivity.
%discuss what drives the high-$\Tc$ $s$-wave superconductivity next to the Mott insulator. 

In the following, we clarify the basic electronic and phonon properties of the fullerides by looking at the realistic parameters 
in the low-energy Hamiltonian in Secs.~\ref{sec_ele} and \ref{sec_int} [the above-described steps (i) and (ii)]. 
Then, we move on to the analysis of the realistic Hamiltonian by the E-DMFT in Secs.~\ref{sec_phase}, \ref{sec_MIT}, 
and \ref{sec_SC} [the step (iii)].

%For the details of the downfolding and model-analysis scheme, see Refs.~\cite{nomura_science_advances,nomura_cDFPT,PhysRevB.85.155452}.

\section{Electronic structure of the fullerides}
\label{sec_ele}

\begin{figure}[tbp]
\vspace{0cm}
\begin{center}
\includegraphics[width=0.47\textwidth]{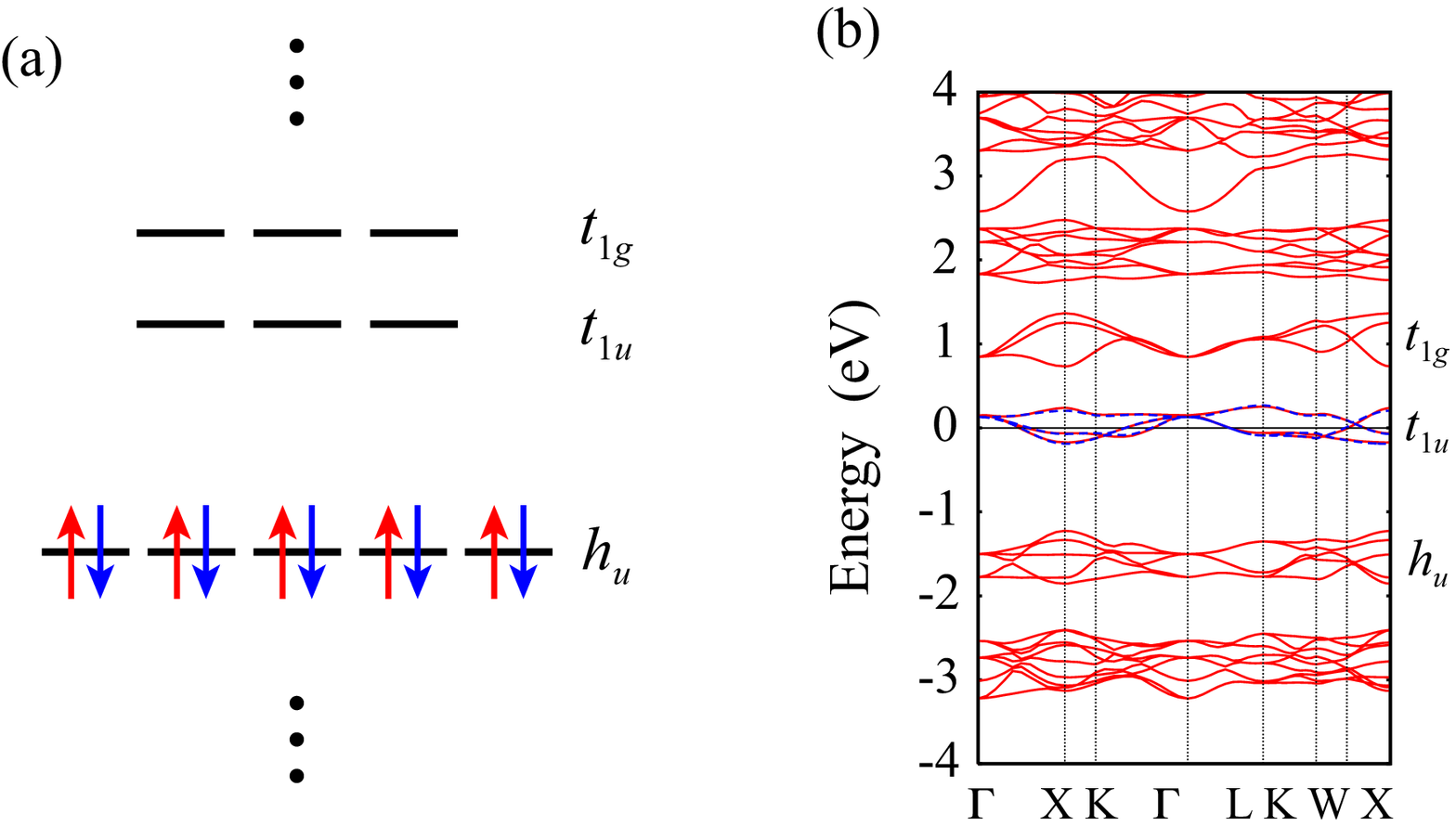}
\caption{(a) Schematic picture for molecular-orbital levels of the \C60 molecule. (b) DFT band dispersion (red) for \Cs3C60 with \VC60 = 762 \AA$^3$. The blue dotted curves show the band dispersion calculated from the one-body part of the low-energy Hamiltonian (see the main text for detail). 
Adapted with permission from \href{http://journals.aps.org/prb/abstract/10.1103/PhysRevB.85.155452}{Nomura {\it et al.}}, Ref.~\cite{PhysRevB.85.155452}. Copyright 2012 by American Physical Society. }
\label{Fig_band}
\end{center}
\end{figure}

Here, we discuss the electronic structure of the fcc \A3C60 systems. 
Since \A3C60 is a molecular solid, the molecular limit is a good starting point for understanding its electronic structure~\cite{ERWIN08111991,PhysRevLett.66.2637} and we expect the bands to arise from the overlap of molecular orbitals.
In Fig.~\ref{Fig_band}(a), we show a schematic picture of the molecular-orbital levels. 
Because of the high symmetry of the \C60 molecule, molecular orbitals often have degeneracy. 
For example, the HOMO orbitals are fivefold degenerate, and the LUMO and LUMO+1 orbitals are threefold degenerate. 
According to their symmetries, these orbitals are called $h_u$, $t_{1u}$, and $t_{1g}$ orbitals, respectively.  

\begin{figure}[tbp]
\vspace{0cm}
\begin{center}
\includegraphics[width=0.45\textwidth]{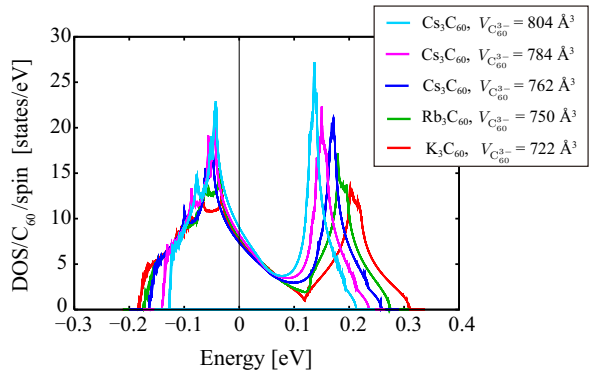}
\caption{DFT density of states for the $t_{1u}$ bands in five different fcc \A3C60 systems. Adapted with permission from 
\href{http://journals.aps.org/prb/abstract/10.1103/PhysRevB.85.155452}{Nomura {\it et al.}}, 
Ref.~\cite{PhysRevB.85.155452}.  Copyright 2012 by American Physical Society.}
\label{Fig_DOS}
\end{center}
\end{figure}

Red curves in Fig.~\ref{Fig_band}(b) show the calculated band structure for fcc \Cs3C60 with \VC60 = 762 \AA$^3$, 
where we have neglected the disorder in the orientations of the \C60 molecules (throughout the paper, we neglect any disorder\footnote{There are several works which study the effect of the merohedral disorder~\cite{nature_K3C60structure,PhysRevB.45.543,PhysRevB.51.5973} on the electronic structure~\cite{PhysRevLett.68.1050,PhysRevB.46.4367,PhysRevB.50.2150}.}).
In solids, the molecular-orbital levels acquire a finite but narrow bandwidth due to the small hoppings between the molecular orbitals~\cite{PhysRevLett.66.2637}.
Because of the narrow bandwidth, each set of bands originating from degenerate molecular orbitals is usually separated from the other molecular bands in energy. 
The intercalated alkali atoms donate electrons into the molecules, hence the LUMO $t_{1u}$ orbitals become half-filled (3 electrons in 3 orbitals).
The doping has little effect on the dispersion of the LUMO bands (i.e., nearly rigid band shift).
The $t_{1u}$ bandwidth can be controlled by applying either chemical or physical pressures and hence changing the lattice constant.  
Figure~\ref{Fig_DOS} shows the DFT density of states of the $t_{1u}$ bands, which clearly shows the expected narrowing of the $t_{1u}$ bandwidth as \VC60 increases. 
 
\begin{figure}[tbp]
\vspace{0cm}
\begin{center}
\includegraphics[width=0.28\textwidth]{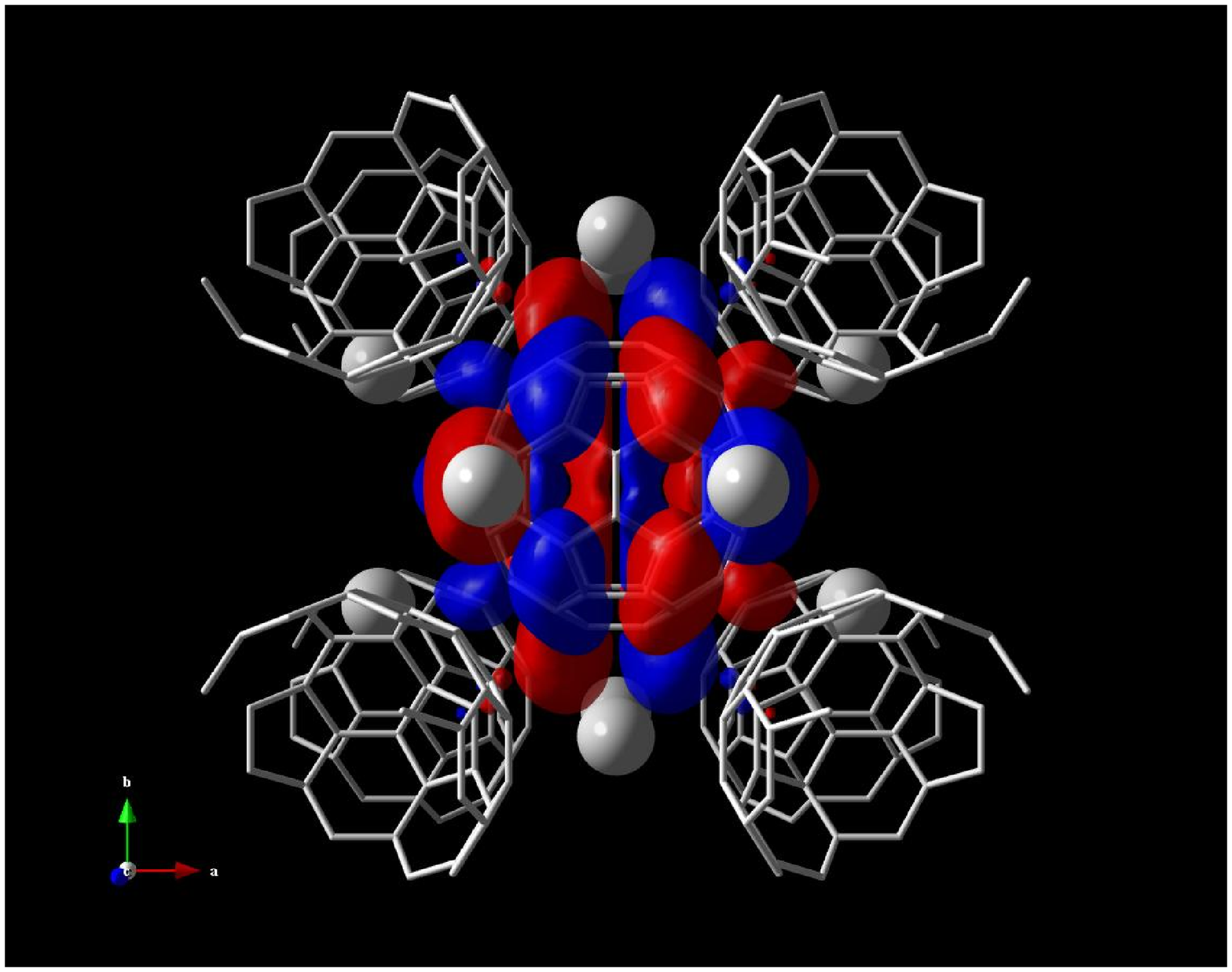}
\caption{One of three maximally localized Wannier orbitals ($p_x$-like orbital) viewed along $z$ direction. 
The positive (negative) isosurfaces of the orbital are depicted by red (blue).
For visibility, we show the Wannier function of A15 \Cs3C60. We note that the shape of the Wannier functions is similar to that of the fcc system. 
Adapted with permission from 
\href{http://journals.aps.org/prb/abstract/10.1103/PhysRevB.85.155452}{Nomura {\it et al.}}, 
Ref.~\cite{PhysRevB.85.155452}.  Copyright 2012 by American Physical Society.}
\label{Fig_Wannier}
\end{center}
\end{figure}

In Ref.~\cite{PhysRevB.85.155452}, to define the basis set for the low-energy Hamiltonian in Eq.~(\ref{H_ele_ph_system}), the maximally localized Wannier orbitals~\cite{PhysRevB.56.12847,PhysRevB.65.035109,RevModPhys.84.1419} were constructed (Fig.~\ref{Fig_Wannier}). 
As expected, the maximally localized Wannier orbitals are very similar to the molecular orbitals, which are centered on 
one
 molecule and are well localized on it. 
We obtain three $t_{1u}$ Wannier orbitals (which can be visualized as $p_x, p_y$, and $p_z$-like orbitals) per molecule. 
By calculating the transfer integral between the molecular orbitals~\cite{PhysRevB.85.155452}, 
we obtain a tight-binding Hamiltonian, which is used as the electronic one-body part $\hat{{\mathcal H}}_{\rm el}$ in Eq~(\ref{H_ele_ph_system}). 
The form of $\hat{{\mathcal H}}_{\rm el}$ is 
\begin{eqnarray} 
 \hspace{0.8cm}  \hat{{\mathcal H}}_{\rm el} = \sum_{i,j,\sigma} t_{ij} \hat{c}_{i\sigma}^{\dagger} \hat{c}_{j\sigma},  
\end{eqnarray}
where $t_{ij}$ is the hopping parameter with $i,j$ being the composite index for the site and orbital. 
Here, each site corresponds to each molecule. 
$ \hat{c}_{i\sigma}^{\dagger}$ ($ \hat{c}_{i\sigma}$) is a creation (annihilation) operator for the electron characterized by the composite index $i$ and the spin $\sigma$. 
Since the three orbitals are degenerate, the double counting correction needed in the DFT+DMFT scheme
becomes a constant shift for all the orbitals, which can be absorbed in the chemical potential. 
The blue-dotted curves in Fig.~\ref{Fig_band}(b) are the dispersion of the $t_{1u}$ bands calculated from $\hat{{\mathcal H}}_{\rm el}$, which reproduces the DFT dispersion. 
It means that the one-body part $\hat{{\mathcal H}}_{\rm el}$ well describes the realistic hopping structure in the fullerides.

\section{Unusual intramolecular interactions}
\label{sec_int}

\begin{figure}[tbp]
\vspace{0cm}
\begin{center}
\includegraphics[width=0.4\textwidth]{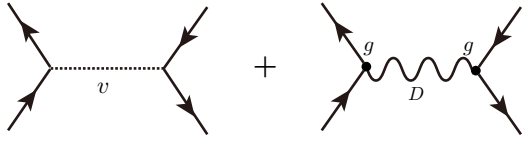}
\caption{Schematic picture which shows that the effective interaction is given by the sum of the Coulomb interaction $v$ 
and the retarded phonon-mediated interaction $g^2D$ with the electron-phonon coupling $g$ and the phonon propagator $D$.  The solid lines denote the electron propagators.}
\label{Fig_interaction}
\end{center}
\end{figure}

\begin{figure}[tbp]
\vspace{0.2cm}
\begin{center}
\includegraphics[width=0.5\textwidth]{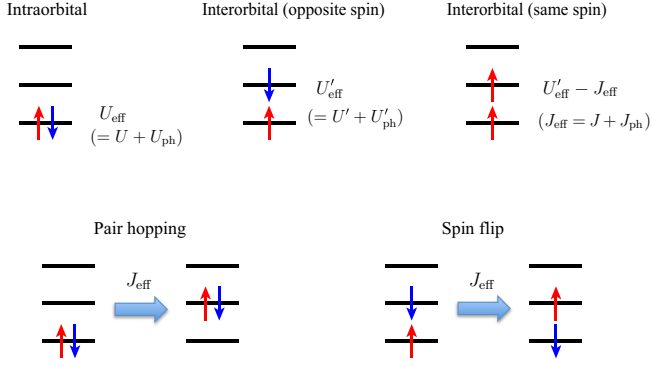}
\caption{Schematic picture for various types of the intramolecular interactions. 
Here, the phonon-mediated interactions are depicted as if they were instantaneous interactions. 
In reality, they are {\it retarded} interactions. }
\label{Fig_form_of_int}
\end{center}
\end{figure}

Next, we turn to the effective interaction between the electrons, which is to be used in the E-DMFT calculation.
This is a very important quantity because the intramolecular interaction dominates the local dynamics of the electrons. 
%the structure of the intramolecular interaction determines the ground state in the atomic limit, which generally helps the understanding of the strongly correlated materials.   
The effective interaction is given by the sum of the repulsive Coulomb and attractive phonon-mediated interactions (see Fig.~\ref{Fig_interaction} for a schematic picture). 
The former and the latter are calculated by the constrained random phase approximation (cRPA)~\cite{PhysRevB.70.195104}, and the constrained density-functional perturbation theory (cDFPT)~\cite{nomura_cDFPT,PhysRevLett.112.027002}, respectively.  
Here, it will be worth emphasizing that the fullerides are multi-orbital systems, which accommodate various types of interactions including density-density type interactions such as the Hubbard $U$, and the non-density type interactions such as the pair-hopping and spin-flip interactions (Fig.~\ref{Fig_form_of_int}).  

In the following, we will show that the fullerides are strongly correlated materials because the intramolecular Hubbard interaction ($\sim 1$ eV) is larger than the low-energy $t_{1u}$ bandwidth ($\sim 0.5$ eV)~\cite{PhysRevB.85.155452}. 
On the other hand, Hund's coupling is found to be very small ($\sim 34$ meV)~\cite{PhysRevB.85.155452}.
This small positive exchange interaction is overcome by a negative contribution from the coupling with the Jahn-Teller phonons, which leads to an effectively negative total exchange interaction~\cite{nomura_science_advances,nomura_cDFPT}, as anticipated on the basis of more phenomenological estimates~\cite{Capone28062002,RevModPhys.81.943}.
%turns to be negative by the contribution from the Jahn-Teller phonons~\cite{nomura_science_advances,nomura_cDFPT}.  
In this section, we start from the Coulomb contribution to the effective interaction in Sec.~\ref{sec_Coulomb}. 
Then, in Sec.~\ref{sec_phonon}, we discuss the phonon contribution. 
Finally, we investigate the total effective interaction in Sec.~\ref{sec_eff_int}.

\subsection{Coulomb interactions} 
\label{sec_Coulomb}

\subsubsection{Formulation}  \hspace{1cm}\vspace{0.25cm} \\
\indent
Using the Wannier basis, we define the Coulomb interaction part  
%$\hat{{\mathcal H}}_{\rm el\mathchar`-el}$ 
\begin{eqnarray}
\hspace{0.8cm} 
\hat{{\mathcal H}}_{\rm el\mathchar`-el} =   \sum_{i,j,k,l} \sum_{\sigma,\sigma'} V_{ij,kl} \ \! \hat{c}_{i\sigma}^{\dagger}  
\hat{c}_{l\sigma'}^{\dagger} 
\hat{c}_{k\sigma'}
\hat{c}_{j\sigma} 
\label{Eq_Coulomb}
\end{eqnarray}
in the low-energy Hamiltonian in Eq.~(\ref{H_ele_ph_system}). 
The Coulomb interaction parameters $V_{ij,kl}$ are calculated by the cRPA~\cite{PhysRevB.70.195104}. 
The cRPA provides {\it partially} screened Coulomb interactions which are screened only by the polarization processes involving  the high-energy bands (the bands other than the $t_{1u}$ bands).
The partially screened interactions can be considered as effective Coulomb interactions within the $t_{1u}$ manifold~\cite{PhysRevB.70.195104}.  
Because the screening processes within the low-energy subspace are considered by the E-DMFT, they are excluded in the cRPA calculation to avoid the double counting of them.

\begin{table}
\caption{\label{tab_cRPA}
The cRPA interaction parameters taken from Ref.~\cite{PhysRevB.85.155452}. 
The values in the parentheses after the material denote \VC60 in \AA$^3$.
$U$, $U'$, and $J$ are intramolecular interactions (see Fig.~\ref{Fig_form_of_int} for their definition), 
for which the relation $U' \sim U-2J$ holds well. 
There is no orbital dependence in $U$, $U'$, and $J$. 
$V$ is the nearest neighbor intermolecular interaction. 
For comparison, the DFT bandwidth $W$ of the $t_{1u}$ bands is also listed.  
} 
\begin{indented}
\lineup
\item[]
\begin{tabular}{l  @{\ } l   @{\  \ \ \  \  }  c@{\  \ \ \    }  c @{\ \ \   }  c @{\ \ \  }  c    @{\ \ \   \ }  c }
\br
%     \multicolumn{2}{l}{\multirow{2}{*}{\hspace{0.6cm}material}}   
    \multicolumn{2}{c}{material}    &  $U$  &  $U'$  &  $J$ & $ V$ & $W$ \\        
    \multicolumn{2}{c}{(fcc structure)}                         & [eV]  &   [eV]  &   [meV]     &  [eV]   &   [eV]     \\   
\mr
 \vspace{0.05cm}   \    \K3C60     &(722)                &   0.82     &  0.76  &  31   & 0.24-0.25  &  0.50 \\
 \vspace{0.05cm}   \   \Rb3C60   & (750)               &   0.92      &  0.85    & 34 &  0.26-0.27 &  0.45    \\
  \vspace{0.05cm}  \   \Cs3C60   & (762)                 &   0.94    &   0.87    &35 & 0.27-0.28  &  0.43  \\
  \vspace{0.05cm}  \   \Cs3C60   & (784)               &   1.02     &   0.94   & 35 & 0.28-0.29   &  0.38   \\
 \vspace{0.00cm}   \   \Cs3C60   & (804)              &   1.07     &   1.00  &  36 & 0.30   &  0.34     \\ 
\br
\end{tabular}
\end{indented}
\end{table}

\subsubsection{Results} \label{sec_cRPA_result}
\hspace{1cm}  \vspace{0.25cm} \\ 
\indent
Table~\ref{tab_cRPA} lists the cRPA interaction parameters.   
The intramolecular Hubbard $U$ is about 1 eV, which is larger than the $t_{1u}$ bandwidth $W$.
Since the ratio $U/W$ exceeds 1,  the alkali-doped fullerides can be regarded as strongly-correlated materials.  
The previous estimates of $U$ in the literature give $U \sim 1$-$1.5$ eV~\cite{PhysRevLett.68.3924,PhysRevB.46.13584,PhysRevB.46.13647,PhysRevB.48.18296}. 
Compared to them, the cRPA values are slightly small. 
%While the material dependence in
Despite that the size of the maximally localized Wannier orbitals depends on materials only weakly~\cite{PhysRevB.85.155452}, 
%we see a nonnegligible 
the material dependence in the Hubbard $U$ is nonnegligible.
This is because the screening strength by the high-energy bands depends on materials~\cite{PhysRevB.85.155452}.
While $U$ is larger than $W$, Hund's coupling $J$ is small $\sim 34$ meV. 
This is explicable by the fact that the Wannier orbitals spread over the molecules, which makes the exchange Coulomb matrix element small. 

The cRPA Coulomb interactions have a long-range tail proportional to $1/r$ with respect to the distance $r$ between the centers of the Wannier orbitals~\cite{PhysRevB.85.155452}, whose nearest-neighbor part $V$ is $\sim 0.25$-$0.3$ eV. 
The offsite Coulomb interactions effectively reduce the size of the onsite Coulomb interaction~\cite{PhysRevB.86.085117}.  
This effect is taken into account by the E-DMFT. 
%We find that the effect of 
While the offsite Coulomb interactions quantitatively
%on the superconductivity and the metal-insulator transition
shift the phase boundaries between e.g., the metal and the insulator,  
they do not play an essential role in driving the superconductivity.  
We find that it is the form of onsite (intramolecular) interactions that is important. 
%we argue that the form of the interactions holds a key 
%quantitatively, and that it does not give a qualitative difference.  
In \ref{app_off}, we discuss this point in more detail.

\subsection{Phonon-mediated interactions} 
\label{sec_phonon}

\subsubsection{Formulation}  \hspace{1cm}\vspace{0.25cm} \\
\indent
The electron-phonon coupling  $\hat{{\mathcal H}}_{\rm el\mathchar`-ph}$ and phonon one-body 
 $\hat{{\mathcal H}}_{\rm ph}$ terms in the low-energy Hamiltonian in  Eq.~(\ref{H_ele_ph_system})  are 
 written as 
 \begin{eqnarray}
 \label{Eq_elph}
 \hspace{0.8cm} \hat{{\mathcal H}}_{\rm el\mathchar`-ph} = \sum_{i,j,\sigma}\sum_{\nu}  \ \! g_{ij}^{\nu}  \ \! \bigl( \hat{c}^{\dagger}_{i \sigma} \hat{c}_{j\sigma}  - \langle  \hat{c}^{\dagger}_{i \sigma} \hat{c}_{j\sigma}    \rangle  \bigr)\ \! \hat{x}_{\nu}
 \end{eqnarray}
 with the displacement operator $\hat{x}_{\nu} = \hat{b}_{\nu}^\dagger + \hat{b}_{\nu}$, and 
 \begin{eqnarray} 
\hspace{0.8cm} \hat{{\mathcal H}}_{\rm ph} = \sum_{\nu}  \omega_{\nu}    \hat{b}_{\nu}^\dagger \hat{b}_{\nu}, 
 \end{eqnarray}
respectively. 
Here, $\hat{b}_{\nu}^\dagger $ ($\hat{b}_{\nu}$) is a creation (annihilation) operator of the phonon and 
$\nu$ is a composite index for the momentum and the branch.
In solids, the electron-phonon coupling $g_{ij}^{\nu}$ and the phonon frequency $\omega_{\nu}$ 
are subject to renormalization by the electrons: 
The electron-phonon coupling is screened by the electronic polarization; the phonons are dressed by electrons, which results in phonon softening.
% of the phonon frequencies. 
As in the case of the Coulomb interaction parameters in Eq.~(\ref{Eq_Coulomb}), in order to avoid the double counting of the 
renormalization, $ g_{ij}^{\nu} $ and $\omega_{\nu}$ in the effective low-energy Hamiltonian should not include the renormalization effect originating from the low-energy electrons~\cite{nomura_cDFPT,PhysRevLett.112.027002,PhysRevB.90.115435}.
The recently developed cDFPT~\cite{nomura_cDFPT,PhysRevLett.112.027002}
  calculates such partially-renormalized phonon quantities by taking into account only the renormalization processes involving the high-energy electrons.  
  It is an extension of the DFPT~\cite{PhysRevB.43.7231,PhysRevB.51.6773,PhysRevB.60.11427,RevModPhys.73.515}, which is a well-established {\it ab initio} scheme to calculate the phonon properties in solids. 
  The expectation value $\langle  \hat{c}_{i \sigma}^\dagger \hat{c}_{j\sigma} \rangle$ in Eq.~(\ref{Eq_elph}) is calculated at the DFT level, where the subtraction of $\langle  \hat{c}_{i \sigma}^\dagger \hat{c}_{j\sigma}    \rangle$  corresponds to the double-counting correction with respect to the equilibrium positions of the ions~\cite{nomura_cDFPT}.

In the action corresponding to the low-energy Hamiltonian in Eq.~(\ref{H_ele_ph_system}), the phonon fields are at most quadratic. 
Therefore, we can analytically integrate out the phonon degrees of freedom without introducing any approximation, which leads to 
%Integrating out the phonon degrees of freedom, we obtain 
an effective action involving only the electronic degrees of freedom.
%Note that we do not introduce any approximation with respect to this integration. 
In this effective action,
the effective interaction between the electrons is given by the sum of the Coulomb interactions and 
the {\it retarded} phonon-mediated interactions (Fig.~\ref{Fig_interaction}).   
This retarded phonon-mediated interactions $V^{\rm ph}_{ij,kl}(i\Omega_n)$ at the Matsubara frequencies ($\Omega_{n} = 2 \pi n T$ with the temperature $T$) are given by~\cite{nomura_cDFPT,PhysRevLett.112.027002} 
\begin{eqnarray}
\hspace{0.8cm}  V^{\rm ph}_{ij,kl} (i\Omega_n)  = -  \sum_{\nu}  \frac{2  g_{ij}^{\nu} g_{kl}^{\nu\ast}  } { \Omega_n^2 + \omega_{\nu}^2}. 
\label{Eq_ph_int}
\end{eqnarray}
An important point here is that the phonon-mediated interaction is given by the sum over the phonon modes. 
We do not assume any particular type of the vibration modes {\it a priori} to study the superconductivity. 
We call the onsite (intramolecular) part of these interactions $U_{\rm ph} (= V^{\rm ph,onsite}_{\alpha\alpha, \alpha\alpha})$, $U'_{\rm ph} (= V^{\rm ph,onsite}_{\alpha\alpha, \beta\beta})$, 
and $J_{\rm ph} (= V^{\rm ph,onsite}_{\alpha\beta, \beta\alpha} = V^{\rm ph,onsite}_{\alpha\beta, \alpha\beta})$, respectively, 
with the orbital indices $\alpha$ and $\beta$.
%the intraorbital, interorbital ($V^{\rm ph,onsite}_{\alpha\alpha, \beta\beta}$), and exchange ($V^{\rm ph,onsite}_{\alpha\beta, \beta\alpha} = V^{\rm ph,onsite}_{\alpha\beta, \alpha\beta}$) components, respectively, .   

\subsubsection{Electron-phonon coupling}\label{sec_el_ph}
 \hspace{1cm}  \vspace{0.25cm} \\ 
\indent
Among 189 ($= 63 \times 3$) phonon branches, it has been argued that the intramolecular lattice vibrations mainly contribute to the total electron-phonon coupling~\cite{RevModPhys.69.575,VARMA15111991,PhysRevLett.68.526}. 
The couplings between the $t_{1u}$ electrons and the other phonon modes such as the intermolecular, alkali-ion, and libration modes have been argued to be small compared to the intramolecular contribution~\cite{RevModPhys.69.575,PhysRevB.48.7651,PhysRevB.46.12088,Pickett_inter,Ebbesen1992163,Burk19942493,PhysRevLett.72.3706}.  
In the molecular limit, by using the group theory, it can be shown that the intramolecular vibrations which couple to $t_{1u}$ electrons are limited to the $A_g$ and $H_g$ modes
(see e.g. Ref.~\cite{book_Jahn-Teller} for a more detailed discussion).
%~\cite{VARMA15111991,PhysRevB.44.12106,PhysRevB.51.3493}.
The $H_g$ vibrations are the Jahn-Teller modes, which induce a split of the $t_{1u}$ energy levels, while they do not change the  
center of the levels~\cite{VARMA15111991,PhysRevB.44.12106,PhysRevB.51.3493}. 
On the other hand, the $A_g$ modes are not of the Jahn-Teller type.  
They couple to the total $t_{1u}$ occupations on the molecule, i.e., they shift the energy levels of each $t_{1u}$ orbital equally.
While the Jahn-Teller modes, which have off-diagonal coupling with respect to the orbital, contribute to $J_{\rm ph}$,  
the non-Jahn-Teller modes, such as the $A_g$ modes, do not. 
We can show that, in the molecular limit, the $H_g$ modes give the intramolecular electron-electron interaction with the form $U_{\rm ph}(i\Omega_n) = - 2 U'_{\rm ph}(i\Omega_n) = 1.5 J_{\rm ph} (i\Omega_n)< 0$~\cite{nomura_cDFPT}.  
The contribution of the $A_g$ modes gives $U_{\rm ph} (i\Omega_n)= U'_{\rm ph} (i\Omega_n)< 0$ and $J_{\rm ph}(i\Omega_n) = 0$.   
  
\subsubsection{Phonon frequencies}\label{sec_freq}
\begin{figure}[tbp]
\vspace{0cm}
\begin{center}
\includegraphics[width=0.42\textwidth]{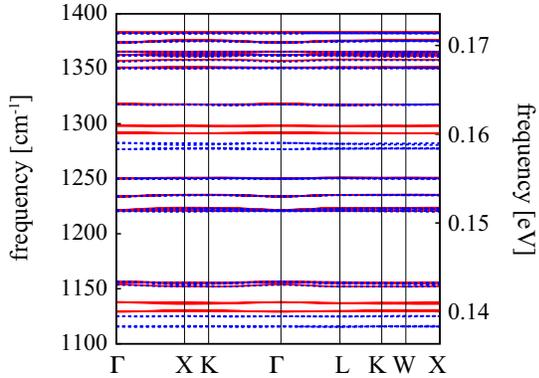}
\caption{Partially-dressed (red) and fully-dressed (blue-dotted) phonon frequencies for fcc \Cs3C60 with \VC60
 = 762 \AA$^3$, calculated by the cDFPT and the DFPT respectively. 
 For visibility, the frequency region is limited from 1100 to 1400 cm$^{-1}$. 
 Reprinted with permission from 
 \href{http://journals.aps.org/prb/abstract/10.1103/PhysRevB.92.245108}{Nomura and Arita}, 
 Ref.~\cite{nomura_cDFPT}. Copyright 2015 by American Physical Society.}
\label{Fig_ph_freq}
\end{center}
\end{figure}

 \hspace{1cm}  \vspace{0.25cm} \\ 
\indent
Here, we look at the partially-dressed phonon frequencies needed to calculate the phonon-mediated interaction in Eq.~(\ref{Eq_ph_int}). 
Figure~\ref{Fig_ph_freq} shows the partially-dressed phonon frequencies (red curves) calculated by the cDFPT
for the frequency range from 1100 to 1400 cm$^{-1}$. 
In the cDFPT calculation, the renormalization of the phonon frequencies coming from the $t_{1u}$ electrons is excluded. 
These partially-dressed frequencies are defined in the low-energy Hamiltonian. 
Thus, they cannot be directly compared to the experimental data. 
To compare with experiments, we need to solve the low-energy Hamiltonian to obtain fully-dressed phonon frequencies.  

In Fig.~\ref{Fig_ph_freq}, for comparison, we also show the fully-dressed phonon frequencies (blue-dotted curves) calculated by the conventional DFPT, where the renormalization effect of the $t_{1u}$ electrons is further incorporated within the DFPT framework.
%In Figure~\ref{Fig_ph_freq},  the frequency range is restricted to 1100-1400 cm$^{-1}$ for visibility. 
The dispersion of both the partially and fully dressed frequencies is tiny, which reflects the intramolecular nature of the modes (almost perfectly described as the Einstein phonons). 
We see that the red curves agree with the blue-dotted curves for the majority of the bands, and that the main difference is in the bands around the frequency of 0.14 and 0.16 eV. 
These two correspond to the two $H_g$ modes out of the eight $H_g$ modes. 
Because of the cubic symmetry, the frequencies of the $H_g$ modes, 
which are fivefold degenerate in the molecular limit, 
are split into threefold- and twofold-degenerate frequencies. 
%In the molecular limit, they are fivefold degenerate.
The $A_g$ modes are located at the energies beyond the range of Fig.~\ref{Fig_ph_freq}. 
%In the frequency range in Fig.~\ref{Fig_ph_freq}, the $A_g$ modes do not exist. 
Experimentally, the two $A_g$ modes are observed at 496 and 1470 cm$^{-1}$ for undoped C$_{60}$~\cite{Bethune1991181}.

The difference between the partially- and fully-dressed phonon frequencies in Fig.~\ref{Fig_ph_freq}
originates from the renormalization effect of the $t_{1u}$ electrons.
% which is considered within the DFPT framework.
Therefore, the frequencies of the $H_g$ modes, which couple to the $t_{1u}$ electrons, are further renormalized from the partially-dressed frequencies in the red curves, while the other modes are not.\footnote{In Fig.~\ref{Fig_ph_freq}, the difference between the partially and fully-dressed phonon frequencies is small. 
This is because the electron-phonon coupling constant $\lambda$ between the $t_{1u}$ electrons and each individual mode is small~\cite{nomura_cDFPT}, while the sum of the contribution from all the modes becomes substantial. }

Because the C-C bonds are rather stiff and the mass of C atom is light, the maximum frequency of the intramolecular phonons
can be rather large, reaching
about 0.2 eV, which is comparable to the $t_{1u}$ bandwidth $\sim 0.5$ eV.   
It indicates that we cannot ignore the vertex corrections, i.e., the Migdal theorem~\cite{Migdal_theorem} does not hold any more.
The low-energy vertex corrections within a molecule can be captured by the E-DMFT, 
while the nonlocal vertex corrections are not.

\subsubsection{Results} \label{sec_ph_result}
\hspace{1cm}  \vspace{0.2cm} \\ 
\indent
Table~\ref{tab_cDFPT} shows the phonon-mediated interactions at zero frequency ($\Omega_n = 0$)
calculated with the partially-renormalized phonon frequencies and electron-phonon interactions~\cite{nomura_science_advances,nomura_cDFPT}. 
The frequency dependence of these interactions is discussed in Sec.~\ref{sec_eff_int}.
The value of the intraorbital interaction $U_{\rm ph}(0)$ lies between $-0.15$ eV and $-0.1$ eV, 
which are small compared to the Hubbard repulsion $U$. 
On the other hand, an unusual situation is realized in the exchange channel: 
The phonon-mediated exchange interaction $J_{\rm ph}(0)$ is about $-51$ meV, and its absolute value is 
larger than the value of Hund's coupling $J \sim 34$ meV. 
As discussed in Sec.~\ref{sec_el_ph}, only the Jahn-Teller phonons contribute to $J_{\rm ph}(0)$. 
Thus, this means that the Jahn-Teller phonons surpass the Coulomb exchange interactions 
and realize an effectively negative exchange interaction.  
However, the size of this negative exchange interaction ($\sim -17$ meV) is small.

\begin{table}
\caption{\label{tab_cDFPT}
The phonon-mediated interactions at zero frequency calculated by the cDFPT. 
The values are taken from Ref.~\cite{nomura_science_advances}. 
The values in the parentheses after the material compositions denote \VC60 in \AA$^3$.
See Fig.~\ref{Fig_form_of_int} for the definition of $U_{\rm ph}$, $U'_{\rm ph}$, and $J_{\rm ph}$.
There is no orbital dependence in $U_{\rm ph}$, $U'_{\rm ph}$, and $J_{\rm ph}$. 
The relation  $U'_{\rm ph} \sim U_{\rm ph}-2J_{\rm ph}$ holds well. 
 } 
\begin{indented}
\lineup
\item[\hspace{0.6cm}]
\begin{tabular}{l  @{\ } l @{\  \ \ \    }   c@{\  \ \ \  \   }  c @{\ \ \  \  \ }  c }
\br
%    \multicolumn{2}{l}{\multirow{2}{*}{\hspace{0.6cm}material}}     &   $U_{\rm ph}(0)$  &  $U'_{\rm ph}(0)$  &  $J_{\rm ph}(0)$ \\ 
     \multicolumn{2}{c}{\hspace{-0.3cm} material}     &   $U_{\rm ph}(0)$  &  $U'_{\rm ph}(0)$  &  $J_{\rm ph}(0)$ \\
     \multicolumn{2}{c}{\hspace{-0.3cm} (fcc structure)}                         &  [meV]  &   [meV]  &   [meV]        \\   
\mr
 \vspace{0.05cm}    \    \K3C60     &(722)   \   \           &   $-152$     &  $-53$  &  $-50$    \\
 \vspace{0.05cm}     \  \Rb3C60   & (750)    \  \            &  $-142$     &  $-42$    & $-51$     \\
  \vspace{0.05cm}   \   \Cs3C60   & (762)    \  \          &  $-114$    &   $-13$    & $-51$   \\
  \vspace{0.05cm}   \   \Cs3C60   & (784)    \  \        &   $-124$     &   $-22$   & $-51$    \\
 \vspace{0.00cm}    \  \Cs3C60   & (804)     \  \       &   $-134$     &   $-31$  &  $-52$    \\ 
\br
\end{tabular}
\end{indented}
\end{table}

\begin{figure}[tbp]
\vspace{0.0cm}
\begin{center}
\includegraphics[width=0.38\textwidth]{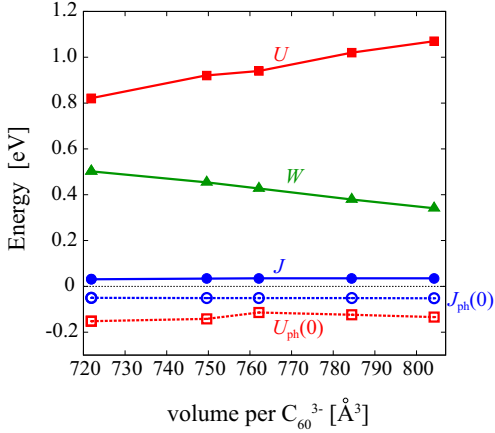}
\caption{ \VC60 dependence of the interaction parameters, $U$, $J$, and $U_{\rm ph}(0)$, and $J_{\rm ph}(0)$, for fcc \A3C60. For comparison, the DFT bandwidth $W$ for the $t_{1u}$ orbitals is also shown.  }
\label{Fig_int_Vdep}
\end{center}
\end{figure}

Figure~\ref{Fig_int_Vdep} summarizes the \VC60 dependence of the Coulomb interactions and the phonon-mediated interactions. 
As \VC60 increases, the Hubbard $U$ increases, while the DFT $t_{1u}$ bandwidth $W$ decreases. 
Thus, the change in $U/W$ is steeper than that expected from the change in the bandwidth.
Hund's coupling $J$ and the phonon-mediated exchange interaction $J_{\rm ph}(0)$ are almost constant throughout the \VC60 range.   In Fig.~\ref{Fig_int_Vdep}, only the phonon attraction $U_{\rm ph}(0)$ has nonmonotonic \VC60 dependence. 
This can be ascribed to the contribution from the alkali-ion vibrations~\cite{nomura_cDFPT}. 
We find that the contribution from the intramolecular modes is nearly \VC60 independent~\cite{nomura_cDFPT}. 
Like the $A_g$ modes, the alkali-ion phonons couple to the total $t_{1u}$ occupations. 
Therefore, they do not contribute to $J_{\rm ph}(0)$. 
The couplings to the alkali modes seem to be nonnegligible, however, one has to pay attention to the fact that they do not include the screening from the $t_{1u}$ electrons.   
When the metallic screening is considered by solving the model, these couplings are efficiently screened and the fully-screened couplings become very small. 
Thus, these alkali modes do not play an important role in the superconductivity. 
On the other hand, the Jahn-Teller-type couplings are poorly screened~\cite{nomura_cDFPT}
and contribute essentially to the superconductivity.

\subsection{Effective interaction between electrons: repulsive Hubbard and negative exchange interactions}
\label{sec_eff_int}

\begin{figure}[tbp]
\vspace{0cm}
\begin{center}
\includegraphics[width=0.38\textwidth]{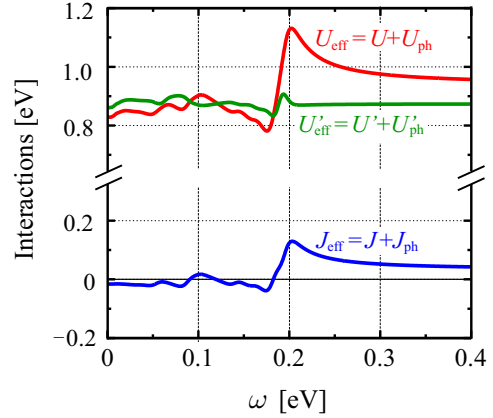}
\caption{Frequency dependence of the real part of the effective intramolecular interactions for fcc \Cs3C60 with \VC60 = 762 \AA$^3$.  
The data are calculated at $\omega + i \eta$ with  a real frequency $\omega$ and $\eta = 0.01$ eV. }
\label{Fig_int_freq_dep}
\end{center}
\end{figure}

Figure~\ref{Fig_int_freq_dep} shows the real-frequency dependence of the effective interactions between the $t_{1u}$ electrons, which are given by the sum of the Coulomb and the retarded phonon-mediated interactions. 
In the frequency region below $0.2$ eV, the attractions from the phonons work since the frequencies of the intramolecular phonons exist up to $\sim 0.2$ eV. 
Several peak-like structures reflect that several phonon modes with different frequencies contribute to the effective interactions.   
By the phonon contribution, the effective exchange interaction $J_{\rm eff}(\omega)$ becomes negative. 
On the other hand, the density-type interactions $U_{\rm eff}(\omega)$ and $U'_{\rm eff}(\omega)$ are strongly repulsive. 
However, in the region where $J_{\rm eff}(\omega)$ becomes negative, $U'_{\rm eff}(\omega)$ becomes slightly larger than $U_{\rm eff}(\omega)$ because of the relation $U'_{\rm eff}(\omega) \sim U_{\rm eff} (\omega) - 2 J_{\rm eff}(\omega)$. 
In the following sections, we discuss the surprising consequence of this unusual relationship between the interactions $U'_{\rm eff} > U_{\rm eff}$ and $J_{\rm eff} <0$.

\section{Theoretical calculation of phase diagram and $\Tc$ from first principles}
\label{sec_phase}

\subsection{Theoretical phase diagram}
\label{sec_phase_theory}

\begin{figure}[tbp]
\vspace{0cm}
\begin{center}
\includegraphics[width=0.44\textwidth]{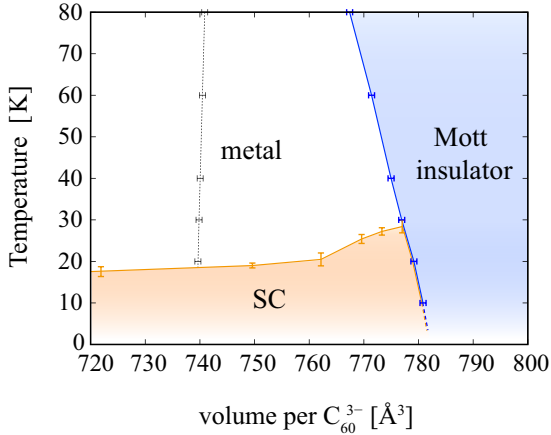}
\caption{Theoretical phase diagram for fcc \A3C60 systems 
obtained by the E-DMFT analysis using the realistic Hamiltonians. 
SC denotes the superconducting phase. 
In between the blue and black-dotted curves, the metallic and insulating solutions coexist in the E-DMFT calculations, either of which is a metastable solution. 
We expect that the first-order transition curve, where the metastable solution changes from 
metal to insulator or vice versa,  is close to the blue curve~\cite{PhysRevB.72.085112}.  
Adapted from 
\href{http://advances.sciencemag.org/content/1/7/e1500568}{Nomura {\it et al.}}, 
Ref.~\cite{nomura_science_advances}. }
\label{Fig_phase_diagram}
\end{center}
\end{figure}

By solving the Hamiltonian in Eq.~(\ref{H_ele_ph_system}) by the E-DMFT,\footnote{In solving the Hamiltonian, we take into account the frequency dependence of the density-density-type interactions~\cite{PhysRevLett.99.146404,PhysRevLett.104.146401,PhysRevB.92.115123}. However, the spin-flip and pair-hopping interactions are assumed to be static to avoid the sign problem in the quantum Monte Carlo solver for the E-DMFT. See Ref.~\cite{nomura_science_advances} for more detail. } 
we derive the theoretical phase diagram~\cite{nomura_science_advances}, 
which is shown in Fig.~\ref{Fig_phase_diagram}. 
%In the single-site E-DMFT calculation, 
Here we  study superconductiviy by 
%The phase boundary between the symmetry-unbroken and symmetry-broken phases are determined by 
directly allowing the symmetry breaking in the E-DMFT calculation. 
Within the single-site E-DMFT calculation for the fcc lattice (non-bipartite lattice), we also allow for ferro-orbital/magnetic order, while we do not consider a possibility of the antiferromagnetic order, which, in the actual fcc lattice, is observed only at very low temperature, significantly smaller than the lowest temperature (10 K) we used in our E-DMFT. 

As a result, we obtain three different phases: the paramagnetic metal, the paramagnetic Mott insulator, and the $s$-wave superconducting phase.  
The $s$-wave superconductivity is characterized by a nonzero superconducting order parameter $\Delta =\sum_{\alpha=1}^{3} \langle c_{\alpha s \downarrow} c_{\alpha s \uparrow}\rangle$, which describes intraorbital Cooper pairs for the $t_{1u}$ electrons. 
Here, $\alpha$ and $s$ are the orbital and site (=molecule) indices, respectively. We omitted a site (=molecule) index in $\Delta$, because $\Delta$ does not depend on a site (the solution is homogenous in space). 
%We find that that $t_{1u}$ electrons form the intraorbital Cooper pairs. 
In the Mott insulating phase, the self-energy diverges and hence the Mott gap opens in the spectral function (the blue-dotted curve in Fig.~\ref{Fig_Aw}). 
Throughout the phase diagram above 10 K, a solution with a ferro-orbital order is not stabilized. 
In the region between the blue curve and the black dotted curve in Fig.~\ref{Fig_phase_diagram}, both metallic and insulating solutions can be stabilized depending on the initial conditions of the E-DMFT calculations.  This suggests that a first-order transition, where the global minimum of the free energy changes from the metallic to insulating solution, should 
take place between
these two curves. According to general entropic arguments, we expect the transition to be close to the blue curve~\cite{PhysRevB.72.085112}.

\subsection{Comparison between theory and experiment}

Comparing the theoretical phase diagram in Fig.~\ref{Fig_phase_diagram} with the experimental one in Fig.~\ref{Fig_phase}(a), 
we find a good agreement between them. 
The theoretical phase diagram reproduces the $s$-wave superconductivity next to the Mott phase. 
Theoretically calculated $\Tc$'s with the maximum of $\sim 28$ K agree with the experimental $\Tc$'s within 10 K difference.  
%It is a remarkable fact 
This agreement is remarkable
since our $\Tc$ calculation does not rely on any empirical parameters: It starts from the DFT calculation using only the information of the crystal structure and all the parameters used in the E-DMFT analysis are calculated from first principles. 

Furthermore, the slope between the paramagnetic metal and insulator is consistent between the theory and experiment. 
As temperature increases, the insulating region expands, which indicates that the insulator has a larger entropy than the metal. 
The position of the metal-insulator boundary is also consistent. 
It is known that the DMFT often overestimates the stability of the metallic phase~\cite{RevModPhys.77.1027,PhysRevLett.101.186403} because it neglects non-local correlation effects and it becomes exact only in the limit of large coordination number. 
However, in the present case, the large coordination number (12) of the fcc lattice makes the DMFT reliable~\cite{nomura_thesis}. 
Furthermore, the \VC60 dependence of the $U/W$ ratio is rather steep so that even if the critical $U/W$ for the metal-insulator transition changes, it does not lead to a drastic change in the critical \VC60.    
We think that the above two factors lie behind the nice quantitative agreement  between the theory and experiment as far as the metal-insulator boundary is concerned.

While we see a good agreement, there are minor discrepancies between the theory and experiment. 
For example, the metal-insulator boundary looks like a crossover in the experiment, while the theory shows a clear first order transition. 
While the reason of the discrepancy is not yet clear, one of the possible reasons is the disorder in the orientations of the buckyballs~\cite{nature_K3C60structure,PhysRevB.45.543,PhysRevB.51.5973}, which is neglected in the calculation. 
Another discrepancy can be seen in the shape of the $\Tc$ curves. 
In the experiment, the $\Tc$ curve shows a dome-like shape, while the theoretical $\Tc$ curve increases toward the Mott transition.  
We come back to this point in Sec.~\ref{sec_SC}.

\section{Property of metal-insulator transition}
\label{sec_MIT}

To gain insight into the superconducting mechanism, we first look at the property of the metal-insulator transition at 40 K.  

\begin{figure}[tbp]
\vspace{0cm}
\begin{center}
\includegraphics[width=0.44\textwidth]{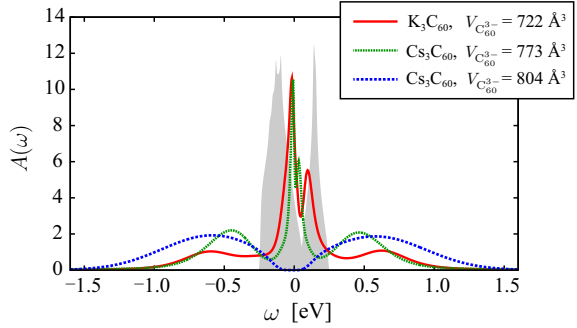}
\caption{Density of states for three different fcc \A3C60 systems, 
obtained by the E-DMFT calculations at 40 K and the subsequent analytic continuation. 
For comparison, the DFT density of states for \K3C60 with \VC60 = 722 \AA$^3$ is shown as the shaded area. 
Adapted from 
\href{http://advances.sciencemag.org/content/1/7/e1500568}{Nomura {\it et al.}}, 
Ref.~\cite{nomura_science_advances}. }
\label{Fig_Aw}
\end{center}
\end{figure}

\subsection{Single-particle spectral function} 

Figure~\ref{Fig_Aw} shows the spectral functions of the $t_{1u}$ bands for three different \A3C60 systems, derived by the analytic continuation based on the maximum entropy method~\cite{PhysRevB.44.6011,Jarrell1996133}. 
The DFT density of states for \K3C60 with \VC60 = 722 \AA$^3$ is also shown for comparison. 
In the metallic phase (red and green curves), as \VC60 increases, i.e., as the correlation strength increases, 
the width of the quasiparticle part becomes narrower and the incoherent peaks become more prominent. 
We note that 
the position of the incoherent part and the renormalization factor of the quasiparticle band for \K3C60 are consistent with the ARPES (angle-resolved photo-emission spectroscopy) measurements for the \K3C60 monolayer~\cite{Yang11042003}.\footnote{The experiment measured a monolayer system. However, the theoretical calculation assumes a bulk system. 
Therefore, we do not compare a band dispersion between the theory and the experiment, and restrict ourselves to the 
comparison of the angle-integrated quantities. }

\subsection{Unusual behaviors across the Mott transition}
\label{sec_Mott_transition}

\begin{figure*}[tbp]
\vspace{0cm}
\begin{center}
\includegraphics[width=0.82\textwidth]{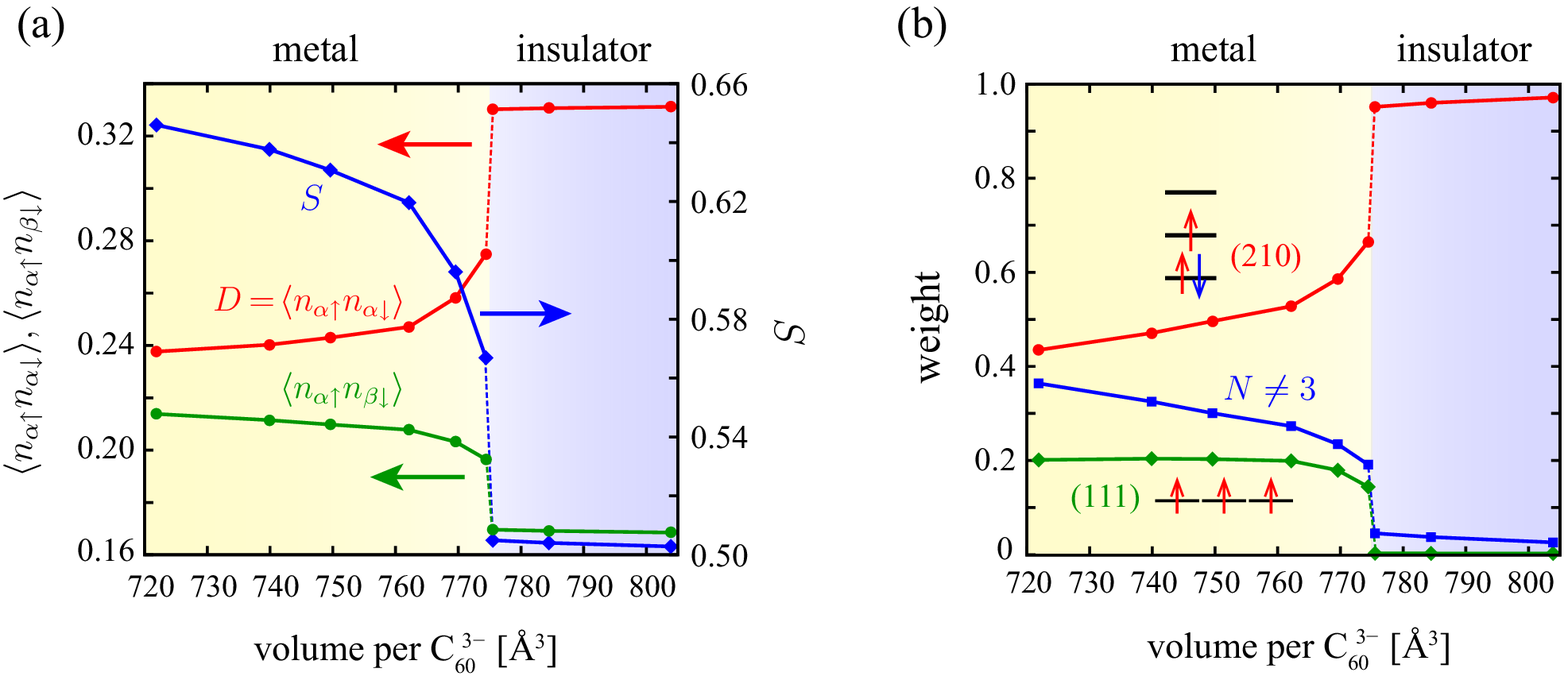}
\caption{(a) \VC60 dependence of the size of spin $S$ per molecule, the double occupancy 
$D=\langle n_{\alpha\uparrow}n_{\alpha\downarrow}\rangle$
on each molecule, and the 
intramolecular
interorbital different-spin correlation function $\langle n_{\alpha\uparrow}n_{\beta\downarrow}\rangle$. 
$\alpha$ and $\beta$ are the orbital indices and we omitted the site (=molecule) index for simplicity. 
(b) \VC60 dependence of weights of intramolecular electronic configurations. 
We also show schematic pictures for the (210) and (111) configurations.
Each (210) configuration, for example a configuration with $\{ n_1, n_2, n_3 \} = \{ 2, 1, 0\}$, 
%lifts an orbital degeneracy as in the figure, 
represents a snapshot of the intramolecular dynamics under the dynamical Jahn-Teller effect.  
When we consider a long-time average, the orbital degeneracy is maintained because all the (210) configurations ($\{ n_1, n_2, n_3 \} = \{ 2, 1, 0\}$, $\{ 0, 2, 1 \}$, $\{ 1, 0, 2  \}$, $\{ 2, 0, 1 \}$, $\{ 1, 2, 0 \}$, and $\{ 0, 1, 2 \}$) have the same weight, 
 i.e., the electrons dynamically fluctuate among (210) configurations (see the main text for more detail).  
In both panels, the results for the metallic solution are shown
in the region where the metallic and insulating solutions coexist in the E-DMFT calculation 
(the region surrounded by the blue and black-dotted curves in Fig.~\ref{Fig_phase_diagram}). 
Adapted from 
\href{http://advances.sciencemag.org/content/1/7/e1500568}{Nomura {\it et al.}}, 
Ref.~\cite{nomura_science_advances}}
\label{Fig_MIT}
\end{center}
\end{figure*}

Figure~\ref{Fig_MIT}(a) shows the dependence of several observables on  \VC60  calculated at 40 K~\cite{nomura_science_advances}.
Here, we focus on the metallic solution in the coexistence region 
(the region delimited by the blue and black-dotted curves in Fig.~\ref{Fig_phase_diagram}). 
The blue curve shows the size of spin $S$ per molecule, which decreases as \VC60 increases. 
In the insulating phase, the value of $S$ becomes $\sim 0.5$, i.e., the low-spin state with $S=1/2$ is realized. 
This is because an effectively negative exchange interaction is realized (Sec.~\ref{sec_eff_int}), which favors the low-spin state 
rather than the high-spin state.  

Another interesting observation is that the double occupancy  $D = \langle n_{\alpha\uparrow}n_{\alpha\downarrow}\rangle$ 
on each molecule
increases toward the Mott transition.
This is in contrast with the intuition and the behavior of the single-band Hubbard model and of multiorbital models with the standard exchange interaction, 
where $D$ is suppressed by the repulsive Hubbard $U$. 
%where $D$ decreases as the correlation strength increases.  
While the fullerides also have a strongly repulsive Hubbard $U$,  the interorbital repulsion $U'$ is effectively larger than the Hubbard $U$ (Sec.~\ref{sec_eff_int}). 
Therefore  the electrons (3 electrons per molecule on average) prefer to sit in the same orbital despite they pay an energy cost of $U$~\cite{PhysRevLett.90.167006}, which is clearly seen in the inequality $\langle n_{\alpha \uparrow} n_{\alpha \downarrow} \rangle >  \langle n_{\alpha \uparrow} n_{\beta \downarrow} \rangle$ in Fig.~\ref{Fig_MIT}(a).

We can gain more insight by looking at the histogram of the weight of the intramolecular electronic configurations
[Fig.~\ref{Fig_MIT}(b)].
The histogram shows which intramolecular electronic state is dominantly realized on the molecule within the E-DMFT~\cite{PhysRevB.75.155113}.  
It is obtained by the continuous-time quantum Monte Carlo simulation~\cite{RevModPhys.83.349,PhysRevLett.97.076405} of the E-DMFT impurity problem, which consists of the single correlated molecule and the dynamical bath. 
%The result is shown in 

In Fig.~\ref{Fig_MIT}(b), ``(210)" denotes the set of the half-filled configurations in which one orbital is doubly occupied and the third electron occupies another orbital,  namely the configurations where $\{ n_1, n_2, n_3 \} = \{ 2, 1, 0\}$, $\{ 0, 2, 1 \}$, $\{ 1, 0, 2  \}$, $\{ 2, 0, 1 \}$, $\{ 1, 2, 0 \}$, and $\{ 0, 1, 2 \}$. 
``(111)" denotes the set of the half-filled configurations with equal occupations on each orbital, i.e., $\{ n_1, n_2, n_3 \} = \{ 1, 1, 1\}$. 
``$N \neq 3$" indicates the total weight of all the configurations away from half filling. 

In the non-interacting limit at half filling, all the 64 intramolecular electronic configurations have equal weight. Then, the weights for 
the (210), (111), and $N \neq 3$ configurations are 0.1875, 0.125, and 0.6875, respectively.
In the presence of correlation effects, the (210) configurations acquire the largest weight. 
This is again because of the unusual molecular interactions with $U'_{\rm eff} > U_{\rm eff}$ and $J_{\rm eff} <0$, which prefer 
the $(210)$-type configurations to the other configurations. 
The increase of the weight of the (210) configurations explains the increase of $D$ and the decrease of $S$ with the increase of \VC60. 

In the metallic phase, however, there exist charge fluctuations because the non-half-filled ($N \! \neq \! 3$) configurations have a nonnegligible weight. 
The non-half-filled ($N \! \neq \! 3$) configurations gradually lose their weight toward the Mott transition. 
In the Mott insulating phase, we see that $N\neq 3$ weight becomes tiny. 
It indicates that the charge degrees of freedom are frozen and the filling on each molecule is nearly fixed at half filling, 
which is nothing but the Mott physics induced by the strongly repulsive $U_{\rm eff}$~\cite{RevModPhys.81.943,Capone28062002}. 

However, the orbital and spin degrees of freedom are active even when the charge degrees of freedom are frozen and the balance of the interaction favors low-spin configurations~\cite{Capone28062002}. 
As we discussed in Sec.~\ref{sec_phase_theory}, the Mott insulating phase has no ferro-orbital/spin order.  
Thus, all the (210) configurations, which are dominant in the Mott phase, are degenerate, offering a room for the orbital and spin fluctuations~\cite{RevModPhys.81.943,Capone28062002}.  
The absence of the orbital order is consistent with the experimentally observed dynamical Jahn-Teller effect~\cite{IR_ncom,1742-6596-428-1-012002}: each (210) configuration can be regarded as a snapshot state where the orbital degeneracy is dynamically lifted, 
however, if we take a long-time average, the orbital degeneracy is recovered. 
We note that the orbital degeneracy plays a role of increasing the critical $U$ for the Mott transition~\cite{PhysRevB.54.R11026,PhysRevB.56.1146}, with stabilizing the metallic/superconducting state against the Mott phase.

\section{Superconducting mechanism}
\label{sec_SC}

Finally, we discuss the superconducting mechanism~\cite{nomura_science_advances}. In order to identify how the different interaction terms give rise to the superconducting state, we check the stability of our superconducting solutions derived from the low-energy Hamiltonian in Eq.~(\ref{H_ele_ph_system}) against the change in the parameters of the Hamiltonian. The results are reported in 
Table~\ref{tab_SC_stability}. 
We see that the superconducting solution becomes unstable when the pair-hopping interaction becomes zero or $U_{\rm eff}$ becomes larger than $U'_{\rm eff}$, while the spin-flip term can be set to zero without destroying superconductivity.
This suggests that the pair-hopping interaction and the relation $U'_{\rm eff} > U_{\rm eff}$ are essential to the superconductivity. 

As discussed in Sec.~\ref{sec_Mott_transition}, $U'_{\rm eff} > U_{\rm eff}$ and associated $J_{\rm eff} < 0$ 
favor the (210) low-spin configurations. 
Accordingly, the pairs of the up- and down-spin electrons with the same orbital character reside on the same molecules, 
which enables the system to form an $s$-wave order parameter with a local pairing amplitude  even in the presence of the strong local Coulomb repulsion.

\begin{table}
\caption{\label{tab_SC_stability}
Stability of the E-DMFT superconducting (SC) solutions at the temperature $T=$ 10 K.
We artificially change the interaction parameters for the E-DMFT from the {\it ab initio} values. 
Then we restart the E-DMFT calculation using the SC solution of the realistic Hamiltonian as an initial guess
and examine whether the SC solution survives or not. 
We try three types of change: (i) We put the pair-hopping interaction to be zero. 
(ii) We put the spin-flip interaction to be zero. (iii) We set $U'_{\rm ph}(i\Omega_n)$ to be equal to $U_{\rm ph}(i \Omega_n)$ 
(which increases the attractions for interorbital channel) so that $U'_{\rm eff}(i\Omega_n) < U_{\rm eff}(i\Omega_n)$ holds for all the boson Matsubara frequency $\Omega_n$. 
We keep other parameters unchanged in each case.
%For these three changes, the parameters which are not mentioned are unchanged. 
The results shown in the table do not depend on \VC60.     
 } 
\begin{indented}
\lineup
\item[\hspace{0.2cm}]
\begin{tabular}{c @{ \ \ \ \ } c @{ \ \ \ \  } c @{ \ \ \ \  } c}
\br 
realistic & no pair-hopping &  no spin-flip  &   $U'_{\rm eff} < U_{\rm eff} $ \\ 
\mr
 SC & no SC & SC  & no SC \\ 
\br
\end{tabular}
\end{indented}
\end{table}

A crucial point is that the formation of the electron pairs is not simply compatible with the strong local repulsion, but it is actually 
assisted and favored by the strong correlations\cite{Capone28062002} in the present multiorbital system.  
Since the difference between $U'_{\rm eff}$ and $U_{\rm eff}$, or in other words, the size of negative $J_{\rm eff}$ ($\sim -17$ meV) is small compared to the typical kinetic energy scale $\sim 0.5$ eV, in the weakly correlated regime, the effect of $J_{\rm eff}$ is small\cite{RevModPhys.81.943}. 
However, the strongly repulsive $U_{\rm eff}$ suppresses the kinetic energy of the electrons, 
driving the system into a regime where the negative $J_{\rm eff}$ is effectively stronger.
Therefore when the repulsion is increased, the negative $J_{\rm eff}$ is more effective in forming local pairs, which can lead to an enhancement of superconductivity. This is made possible by the fact that the pairing acts in the spin-orbital channel, 
which remains active even in the presence of the strong correlation\cite{Capone28062002}.

The pair-hopping interaction, whose amplitude is again given by $J_{\rm eff}$, is also important, as discussed in several previous works~\cite{RevModPhys.81.943,Capone28062002,PhysRevB.44.10414,PhysRevLett.93.047001,PhysRevB.46.1265,doi:10.1143/JPSJ.69.2615}. 
The generated intraorbital pairs can tunnel into another orbital through the pair-hopping process, which also enhances the superconductivity (the Suhl-Kondo mechanism~\cite{PhysRevLett.3.552,Kondo01011963}). 
In principle, the enhancement of the superconductivity by the Suhl-Kondo mechanism occurs irrespective of the sign of the pair-hopping interactions. 
However, in the present three-orbital system, 
the negative pair-hopping enhances the superconductivity more efficiently than the positive one.
This is because the former prefers the gap function with the same amplitude and sign for every orbital, while the latter favors the sign change in the gap function, generating a frustration.  

%in the three orbital case, the enhancement by the negative pair-hopping interaction, 
%which prefer the gap functions with the same amplitude and the same sign for each orbital, 
%is more efficient than the positive case, 
%which favors the sign change of the gap functions, inducing a frustration in the three orbital model.  

With the above considerations, we conclude that 
the crucial factors for the $s$-wave superconductivity in the alkali-doped fullerides are~\cite{nomura_science_advances} 
\begin{itemize}
\item the formation of the intraorbital electron pairs on molecules induced by the unusual intramolecular interactions with $U'_{\rm eff} > U_{\rm eff}$, which becomes more efficient under the strong correlation
\end{itemize}
and 
\begin{itemize}
\item the interorbital tunneling of the electron pairs through the pair-hopping interaction. 
\end{itemize}
%are the crucial factors for the $s$-wave superconductivity in the alkali-doped fullerides~\cite{nomura_science_advances}. 
Since the unusual relations $J_{\rm eff} < 0 $ and $U'_{\rm eff} > U_{\rm eff}$ are  a consequence of the coupling of the electrons with the Jahn-Teller phonons,
we confirm that the phonons are necessary to drive the $s$-wave superconductivity in these materials. 
However, thanks to the unusual form of the multi-orbital interactions, the strong correlations surprisingly cooperate with the phonons by assisting and favoring the formation of the intraorbital electron pairs. This fact marks a sharp contrast with the conventional phonon mechanism, where the phonons and the Coulomb interactions compete with each other, especially when both interactions are local and for half-filled bands\cite{Sangiovanni}.

The picture emerging from our fully {\it ab-initio} theory is consistent with the previous model studies on the negative-$J$ multiorbital Hubbard model~\cite{RevModPhys.81.943,PhysRevLett.86.5361,Capone28062002,PhysRevLett.93.047001}, where $J$ in this model can be considered as  
$J_{\rm eff}$ in the {\it ab initio} low-energy Hamiltonian. 
In these studies it was argued that when the size of the negative $J$ is small, a strongly-correlated superconductivity emerges in the vicinity of the Mott transition: It benefits from the strong correlation and is distinct from the BCS-type superconductivity. 
Close to the Mott transition, the charge fluctuations are suppressed by the strong Hubbard interaction
and the quasiparticle bandwidth is renormalized as $ZW$ with $Z \ll 1$ being the quasiparticle weight.
% which renormalizes the kinetic energy.
Accordingly, the quasiparticles effectively feel renormalized Hubbard interaction $ZU$.
On the other hand, $J$ works on the spin and orbital degrees of freedom, which are still active even when the charge
degrees of freedom are frozen (Sec.~\ref{sec_Mott_transition}).
Since $J$ does not directly see a freezing of the charge degrees of freedom, $J$ is not renormalized by the correlation.  
As a result, the heavy quasiparticles feel the renormalized repulsion $ZU$ and the unrenormalized attraction $J$. 
This is why the strong correlation helps the $s$-wave superconductivity: It strongly reduces the residual repulsive interaction between the heavy quasiparticles, but it does not affect the small bare attractive interaction arising from electron-phonon coupling. Therefore sufficiently close to the Mott transition, the overall effective interaction turns into an attraction.

Finally, we discuss 
the origin of the dome shaped $T_c$ in the experimental phase diagram, 
which is not seen in our theoretical result: The calculation gives an increase of the critical temperature as the transition to the Mott insulator is approached.
%the absence of the clear dome shape in the theoretical $\Tc$ curve, 
%while the experiment observes the downturn of $\Tc$ toward the metal-insulator transition.  
As shown in the study on the negative-$J$ multiorbital Hubbard model~\cite{RevModPhys.81.943,PhysRevLett.86.5361,Capone28062002,PhysRevLett.93.047001}, the superconductivity is eventually suppressed toward the 
Mott transition when $Z$ becomes tiny. 
However, in the calculation for the realistic Hamiltonian at finite temperatures, the transition to the Mott insulating state
is of first order and the $\Tc$ curve is cut before we see a downturn of it.  
Indeed at $T = 10$ K, we observe a dome shape in the superconducting order parameter $\Delta$ if we follow the superconducting solution
($\Delta$ = 0.028, 0.031, 0.033, and 0.029 at 10 K for \VC60 = 762.2, 767.7, 773.3, and 778.9 \AA$^3$, respectively).
It
indicates that 
the $\Tc$ dome is hidden by the insulating phase in the current calculation. 
Other factors,  which are not considered in the calculation such as the effect of the merohedral disorder~\cite{nature_K3C60structure,PhysRevB.45.543,PhysRevB.51.5973} and the non-local fluctuations, might also play a role in the shape of the experimental $\Tc$ curve, which remain to be investigated.

\section{Conclusion and future perspective}
\label{sec_summary}

\subsection{Summary of the review}
We have reviewed the properties of alkali-doped fulleride superconductivity, starting from the basic electronic structure of the fcc \A3C60 systems. 
The band structure of \A3C60 fullerides is well described by a picture in terms of molecular orbital levels connected by relatively small hoppings between them. 
The half-filled low-energy $t_{1u}$ bands are the main playground for the superconductivity. 

We have elucidated that the effective intramolecular interactions between the $t_{1u}$ electrons, 
which arise from the combination of the Coulomb repulsion and electron-phonon interactions, 
have an unusual structure characterized by a strongly repulsive Hubbard repulsion and a weakly negative exchange interaction, which leads to an inverted Hund's coupling. 
This unusual situation occurs since the fullerides are the degenerate multi-orbital systems having tiny Hund's coupling and strong coupling to Jahn-Teller modes.   

The realistic Hamiltonian with this form of the intramolecular interaction comprehensively reproduces the experimental phase diagram including the adjacency of the Mott insulator and the $s$-wave superconductivity.
Remarkably, the agreement is not only qualitative but also quantitative; the theoretically calculated $\Tc$ using only 
the information of the crystal structure, agrees with the experimental $\Tc$ within a difference of 10 K.     
The derived Mott insulating phase is characterized by a low-spin ($S=1/2$) state with orbital degeneracy, completely consistent with the experimental observations. 

The analysis on the superconducting mechanism reveals that the high-$\Tc$ $s$-wave superconductivity is driven by an unusual cooperation between the strong correlations and the Jahn-Teller phonons. 
This confirms that the mechanism of the fulleride superconductivity is indeed unconventional and requires strong correlation effects, despite the crucial role of the electron-phonon interaction and the $s$-wave symmetry of the order parameter.

\subsection{Future perspective}

\subsubsection{Study on A15 \Cs3C60}
\hspace{1cm}  \vspace{0.2cm} \\ 
\indent
In this review, we have not discussed the A15 systems and have focused on the fcc systems. 
Through the calculations for the fcc systems, we have found that the dynamics of the electrons within the molecule with unusual form of the interactions is the most important factor to drive the 
$s$-wave superconductivity. 
The building block of the A15 systems is the same \C60 molecules as that in the fcc systems. 
Then, as far as an effectively negative exchange interaction is realized on each molecule, the intraorbital pair-formation and the Suhl-Kondo mechanism, which are related to intramolecular dynamics, will work irrespective of the lattice structure. 
%\rout{We expect that such intramolecular dynamics is not much affected by the change of the lattice structure.}
This explains naturally why the superconducting properties are similar between the A15 and fcc systems in experiment. 
On the other hand, the antiferromagnetism will be strongly affected by the lattice structure: For example, the degree of frustration and the structure of the super-exchange interactions are crucial for the magnetism. 
It will be interesting to apply the present scheme also to the A15 systems and see whether it also comprehensively 
explains their experimental phase diagram. 

\subsubsection{Light-induced superconducting-like state in \K3C60}
\hspace{1cm}  \vspace{0.2cm} \\ 
\indent
Recently, superconducting-like signatures have been reported in the optical spectra 
of  \K3C60 at temperatures higher than 100 K (much higher than the equilibrium critical temperature $\Tc$ = 19 K)
by means of impulsive excitation~\cite{cavalleri_paper} designed to coherently excite the molecular vibration of the fullerene molecules.
This inherently nonequilibrium phenomenon has not yet been understood and is surely a fascinating challenge for theory. 
We believe that the present understanding of the equilibrium superconductivity provides a firm basis for studying the nonequilibrium superconducting-like state.

\ack
We would like to thank K. Nakamura for helpful discussions and for the technical support. 
We acknowledge fruitful discussions with Y. Iwasa Y. Kasahara, K. Prassides, P. Werner, T. Ayral, M. Kim, A. Georges, Y. Murakami, H. Shinaoka, T. Kosugi, N. Parragh, G. Sangiovanni, M. Imada, A. Oshiyama, A. Fujimori, P. Wzietek, H. Alloul, M. Fabrizio, E. Tosatti, and G. Giovannetti. 
%In particular, we thank Y. Iwasa and Y. Kasahara for providing us with the experimental phase diagram. 
%Some of the calculations were performed at the Supercomputer Center, ISSP, University of Tokyo. 
Y.N., S.S., and R.A. were  
supported by Grant-in-Aid for JSPS Fellows (no. 12J08652), 
Grant-in-Aid for Scientific Research (no. 26800179), 
%from JSPS Japan. 
Grant-in-Aid for Scientific Research (no. 15H03696) from Japan Society for the Promotion of Science (JSPS), Japan, respectively. 
M.C. is supported by FP7/European Research Council (ERC) through the Starting Grant SUPERBAD (grant agreement no. 240524) and by the EU-Japan Project LEMSUPER (grant agreement no. 283214).

\appendix
\section{Effect of intermolecular Coulomb interactions}
\label{app_off}

The E-DMFT takes into account the effect of the intermolecular Coulomb interactions
as a dynamical reduction of the intramolecular interactions. 
We find that this dynamical screening is not an essential ingredient in the superconductivity, 
while it quantitatively improves the location of the Mott transition in Fig.~\ref{Fig_phase_diagram}~\cite{nomura_science_advances}. 
Since the orbital dependence of the intermolecular interactions is negligible, 
the resulting dynamical screening on the intramolecular interactions has no orbital dependence. 
That is, it equally reduces the intraorbital interaction $U_{\rm eff}$ and the interorbital interaction $U'_{\rm eff}$  
and does not affect the exchange channel $J_{\rm eff}$.
Since the negative $J_{\rm eff}$ is a key for the unconventional physics in the fullerides, 
even without the dynamical screening effect from the intermolecular interactions, 
we obtain a phase diagram qualitatively similar to Fig.~\ref{Fig_phase_diagram}. 
However, since the intermolecular interactions reduce the value of $U_{\rm eff}$ and $U'_{\rm eff}$, they stabilize a metallic solution and shift 
the metal-insulator phase boundary toward a larger \VC60, making the agreement between theory and experiment better. 
\vspace{0.5cm}

\bibliographystyle{iopart-num}
\bibliography{C60_review}

\end{document}